\documentclass[a4paper,12pt]{article}

\usepackage{graphicx}
\usepackage{array}
\usepackage{amsmath}
\usepackage{amssymb}
\usepackage{latexsym}
\usepackage{subfigure}
\usepackage{cite}
\usepackage{color}

 \usepackage[normalem]{ulem}

 \renewcommand\sout{\bgroup \color{red} \ULdepth=-.5ex \ULset}

\date{empty}
\begin{document}
\begin{titlepage}
\null
\begin{flushright}
June, 2017
\end{flushright}
\vskip 2.0cm
\begin{center}
{\Large \bf
Supersymmetry Breaking in Spatially Modulated Vacua
}
\vskip 1.7cm
\normalsize
\renewcommand\thefootnote{\alph{footnote}}

{\large
Muneto Nitta$^{\dagger}$\footnote{nitta(at)phys-h.keio.ac.jp},
Shin Sasaki$^\ddagger$\footnote{shin-s(at)kitasato-u.ac.jp}
and
Ryo Yokokura$^\sharp$\footnote{ryokokur(at)rk.phys.keio.ac.jp}
}
\vskip 0.7cm
{\it
$^\dagger$
Department of Physics, \& Research and Education Center for Natural Sciences, \\
\vskip -0.2cm
Keio University, Hiyoshi 4-1-1, Yokohama, Kanagawa 223-8521, Japan
\vskip 0.1cm
$^\ddagger$
Department of Physics,  Kitasato University, Sagamihara 252-0373, Japan
\vskip 0.1cm
$^\sharp$
Department of Physics, Keio University, Yokohama 223-8522, Japan
}
\vskip 0.5cm
\begin{abstract}
We study spontaneous supersymmetry breaking in spatially modulated
stable or meta-stable vacua in supersymmetric field theories.
Such spatial modulation can be realized in a higher derivative chiral
 model for which vacuum energies are either positive, negative or zero,
depending on the model parameters.
There appears a Nambu-Goldstone boson associated with the spontaneously
 breaking of the translational and $U(1)$ symmetries without the quadratic kinetic
 term and with a quartic derivative term in the modulated direction, and a gapless Higgs mode.
We show that there appears a Goldstino associated with the supersymmetry
 breaking at a meta-stable vacuum,  where energy is positive,
 while it becomes a fermionic ghost in the negative energy vacuum,
 and zero norm state and disappears from the physical sector
 in the zero energy vacuum.

\end{abstract}
\end{center}

\end{titlepage}

\newpage


\newpage
\section{Introduction}
Finding vacua where supersymmetry (SUSY) is spontaneously broken is an
important problem in supersymmetric field theories,
since it is obviously broken if it exists in nature.
The famous examples of spontaneous SUSY breaking include the
O'Raifeartaigh model for chiral superfields
\cite{ORaifeartaigh:1975nky} and supersymmetric gauge theories \cite{Fayet:1974jb},
where the positive energy vacuum is characterized by a constant vacuum expectation value of scalar fields.
Remarkably, it is desirable that these spontaneous SUSY breakings are caused by the
dynamics of models \cite{Witten:1981nf}.
However, the severe constraint by the Witten index \cite{Witten:1982df} makes it hard to construct a
phenomenologically viable model where dynamical SUSY breaking is possible.
A large amount of efforts has been devoted to construct a model for
the dynamical SUSY breaking.
The constraint of the Witten index can be circumvented if one employs a
local minima, not the global minimum, for the SUSY breaking vacua.
Even though the local minima are
meta-stable false vacua decaying
into the global vacuum in a finite time, they are nevertheless useful candidates of
phenomenologically possible vacua
if the life time of the vacua
is
longer than
that of our
 Universe.
This is the idea of the SUSY breaking in the meta-stable vacua
\cite{Intriligator:2006dd, Intriligator:2007cp}.
It is worthwhile to emphasize that almost all of the SUSY breaking vacua
discussed in the literature respect the translational symmetry in the
relativistic field theories, for which the order parameter of vacua is
constant.

On the other hand, space-time symmetry breakings
have been discussed in a vast literature.
Nonlinear realizations for spontaneously broken 
space-time symmetry were first formulated in Ref.~\cite{Ivanov:1975zq} 
as the so-called inverse Higgs mechanism, 
and corresponding  Nambu-Goldstone (NG) modes were discussed in Ref.~\cite{Brauner:2014aha}. 
Phenomenology of the spontaneous Lorentz symmetry breakings have been
intensively studied in the past \cite{Kostelecky:1988zi,
Bluhm:2004ep, Altschul:2005mu, Kostelecky:2009zr, Bluhm:2007bd}.
The ghost condensation 
\cite{ArkaniHamed:2003uy} also gives an example.
The presence of a brane or soliton also 
breaks translational symmetry perpendicular to the brane 
as well as a Lorentz symmetry tilting the brane.
In this case,
the NG modes associated with the broken symmetries appear
as massless fields in the world-volume theory
\cite{Aganagic:1996nn, Adawi:1998ta, Kaplan:1995cp} (and references in \cite{Simon:2011rw}).
Spontaneous breakings of the (super)Poincar\'e symmetry 
have been also discussed in the context of BPS \cite{Bagger:1996wp,
Kallosh:1997aw, Ivanov:1999fwa} as well as non-BPS branes \cite{Clark:2002bh}.
In addition to spontaneous breakings, there are also studies on 
the explicit Lorentz violations from the view
points of quantum gravity \cite{Horava:2009uw,Sotiriou:2009bx},
massive gravity \cite{Dubovsky:2004sg, Nojiri:2010wj}
and particle physics
\cite{Coleman:1998ti,Kostelecky:2003fs,Colladay:1998fq}.
The explicit Lorentz symmetry violations in SUSY theories 
\cite{Berger:2001rm, Berger:2003ay, Bolokhov:2005cj},
including formal aspects of Lorentz violating SUSY breaking 
\cite{GrootNibbelink:2004za},
have been also discussed.

Among other things, 
it is becoming more important to consider the possibility of
spatially inhomogeneous ground states in
condensed matter physics \cite{Fulde:1964zz,larkin:1964zz}
and QCD \cite{Nakano:2004cd,Nickel:2009ke,Buballa:2014tba}.
For such a kind of ground states, the order parameter is characterized by a spatially varying function 
and several translational symmetries are spontaneously broken there.
We have recently proposed that
such modulation can occur in relativistic field theories \cite{Nitta:2017mgk},
and have found that NG boson appears
as a consequence of spontaneous symmetry breaking
of translational and $U(1)$ symmetries. 
Despite the physical importance of the spatially modulated vacua, 
there have been no studies on such vacua in supersymmetric contexts.
It would be therefore plausible to admit SUSY breaking in spatially inhomogeneous vacua where parts of space-time symmetries in theories are also broken.

In this paper, we study spontaneous SUSY breaking in
modulated vacua where the translational symmetry is broken.
This possibility may open up phenomenologically viable model buildings
based on a new kind of SUSY breaking.
Our model contains a SUSY breaking modulated vacuum in addition to the SUSY preserving vacuum.
The modulated vacuum which we find is either meta-stable with positive vacuum energy,
stable and degenerated with the SUSY preserving vacuum which has zero vacuum energy, or unstable with negative vacuum energy, depending on the model parameters.
In addition to the NG boson associated with spontaneously broken translational symmetry \cite{Nitta:2017mgk},
there appears a massless fermion, a Goldstino,
in any case, as a consequence of the SUSY breaking.
In the case of the positive energy vacuum, the Goldstino propagates with
the correct sign
of the kinetic term
both along
the modulation and the transverse directions.
For the zero energy vacuum, the Goldstino has zero norm and disappears from the spectrum,
even in the presence of the SUSY breaking.
Although the negative energy vacuum is stable in the bosonic sector
\cite{Nitta:2017mgk}, it is unstable in the fermionic sector:
the Goldstino becomes a fermionic ghost in the orthogonal direction,
thereby leading to the instability.
It can have a kinetic term with the wrong sign
even along the modulated direction when the
vacuum energy is negative or zero.
One of the interesting features of our model is that
SUSY is broken even though auxiliary field $F$ does not have a vacuum
expectation value (VEV), unlike usual SUSY breakings.

In order to find such kind of modulated vacua, we introduce
supersymmetric higher derivative chiral models.
From a viewpoint of low-energy effective theories, supersymmetric field
theories generically receive higher derivative corrections. 
Here 
``the higher derivative''
means that terms that contain more than two space-time derivatives.
There are variety of higher derivative supersymmetric chiral models.
We concentrate on models where only the single space-time derivative
acts on fields, like $\partial^m \varphi$. 
In this paper, we never consider terms that contain 
more than two derivatives on one field like $\partial^2 \varphi$
that cannot be removed by partial integrations in the action. 
Terms with this kind of interactions suffer from a potential instability
of systems \cite{Antoniadis:2007xc}.
This instability results in the existence of ghosts and it 
is known as the Ostrogradski instability \cite{Ostrogradski}
\footnote{
There is a way to remove a ghost by gauging
\cite{Fujimori:2016udq}, but we do not consider such a possibility.
}.
It is convenient to employ the off-shell superfield formalism to
construct supersymmetric theories.
One often encounters the so-called auxiliary field problem implying that
the equation of motion for the auxiliary field $F$ cease to be algebraic
\cite{Gates:1995fx}.
Then it is not so easy to write down the on-shell Lagrangians.
This has been seen in various supersymmetric higher derivative models,
such as a supersymmetric WZW term \cite{Nemeschansky:1984cd},
supersymmetric Skyrme models
\cite{Bergshoeff:1984wb}
 and so on.
The supersymmetric higher derivative models free from the auxiliary
field problem has been discussed in various contexts.
For example, higher derivative corrections to the ordinary quadratic
kinetic terms appear in low-energy effective theories of supersymmetric
models \cite{Buchbinder:1994iw, Gomes:2009ev}.
Other examples include the supersymmetric generalization of the
Wess-Zumino-Novikov-Witten (WZNW) term \cite{Gates:2000rp}, the world-volume
action of supersymmetric branes \cite{Rocek:1997hi}, higher derivative
chiral models coupled with supergravities \cite{Farakos:2012qu}, supersymmetric
Skyrme-like models \cite{Adam:2011gc, Gudnason:2015ryh} and an inflation model driven by
supersymmetric higher derivative terms of inflatons
\cite{Sasaki:2012ka}.
Two of the present authors have studied 
BPS states in supersymmetric higher derivative theories
\cite{Nitta:2014pwa} and higher derivative corrections to manifestly
supersymmetric nonlinear realization of the NG multiplet
\cite{Nitta:2014fca}. 
In particular, all possible four derivative terms
free from auxiliary field problem and ghosts
have been classified in Ref.~\cite{Khoury:2010gb},
and
they have been generalized to arbitrary number of derivatives in
Refs.~\cite{Nitta:2014pwa,Nitta:2014fca}, which we use in this paper.

The organization of this paper is as follows.
In Sec.~\ref{sec:model}, we introduce the supersymmetric chiral model with
higher derivative terms that is free from the auxiliary field problem in the bosonic sector.
In Sec.~\ref{sec:modulation}, we focus on a specific model where
spatially modulated ground states are allowed.
Supersymmetry is spontaneously broken in the modulated vacua.
We show that the
modulated vacuum is classified according to the vacuum energy.
In Sec.~\ref{sec:NG}, we discuss the NG modes in the modulated
vacua. We demonstrate that the quadratic kinetic terms of bosonic NG modes
associated with the spontaneously breaking of bosonic symmetries in the
modulated vacua vanish in general. On the other hand,
a Higgs mode,
perpendicular to the NG mode, appears as a massless boson.
For the spontaneous breaking of SUSY, the corresponding
NG mode, {\it i.e.} the Goldstino becomes a ghost when the
vacuum energy is negative while it becomes a zero norm state when the
vacuum has zero energy.
In Sec.~\ref{sec:superpot}, we introduce a superpotential in our model.
Section \ref{sec:summary} is devoted to conclusion and discussions.
The component expansions of the higher derivative parts of the chiral
superfield are found in Appendix A.

\section{Supersymmetric higher derivative model}\label{sec:model}
In this section we introduce the supersymmetric higher derivative model
which is free from the auxiliary field problem in the bosonic sector.
The Lagrangian of the model is given by
\begin{align}
\mathcal{L} &=
\ \int \! d^4 \theta \ K (\Phi^i, \Phi^{\dagger \bar{j}})
+ \frac{1}{16} \int \! d^4 \theta \ \Lambda_{ik\bar{j} \bar{l}} (\Phi, \Phi^{\dagger})
D^{\alpha} \Phi^i D_{\alpha} \Phi^k \bar{D}_{\dot{\alpha}} \Phi^{\dagger
 \bar{j}} \bar{D}^{\dot{\alpha}} \Phi^{\dagger \bar{l}}  \nonumber \\
 &+ \left(\int \! d^2 \theta \ W(\Phi^i) + h.c.\right).
\label{eq:superfield_Lagrangian}
\end{align}
Here $\Phi^i = \varphi^i (y) + \sqrt{2} \theta^{\alpha} \psi_{\alpha}^i
(y) + \theta^2 F^i(y) \ (i = 1, \ldots, N)$ are the four-dimensional
$\mathcal{N} = 1$ chiral superfields in the chiral base $y^m = x^m + i
\theta \sigma^m \bar{\theta} \ (m=0,1,2,3)$
whose component fields are complex scalars $\varphi^i$, Weyl fermions
$\psi^i$ and auxiliary fields $F^i$.
$K$ is a K\"ahler potential, $W$ is a superpotential and $\Lambda_{ik\bar{j}
\bar{l}}$ is a $(2,2)$ K\"ahler tensor whose (anti)holomorphic indices are symmetrized.
We basically follow the conventions and notations of Wess and Bagger \cite{Wess:1992cp}.
The flat metric is given by $\eta_{mn} = \mathrm{diag} (-1,1,1,1)$.
The first and the third terms in \eqref{eq:superfield_Lagrangian} are
the ordinary kinetic and potential
terms of supersymmetric chiral models while the second term provides higher
derivative terms.
A specific property of the second term is
that the purely bosonic components included in there saturate the Grassmann coordinates:
\begin{align}
& \frac{1}{16} D^{\alpha} \Phi^i D_{\alpha} \Phi^k \bar{D}_{\dot{\alpha}}
 \bar{\Phi}^{\dagger \bar{j}} \bar{D}^{\dot{\alpha}} \Phi^{\dagger
 \bar{l}}  \nonumber\\
 &= \theta^2 \bar{\theta}^2
\left[
(\partial_m \varphi^i \partial^i \varphi^k) (\partial_n
 \bar{\varphi}^{\bar{j}} \partial^n \bar{\varphi}^{\bar{l}}) - 2
 \partial_m \varphi^i F^k \partial^m \bar{\varphi}^{\bar{j}}
 \bar{F}^{\bar{l}} + F^i \bar{F}^{\bar{j}} F^k \bar{F}^{\bar{l}}
\right] + I_{\text{f}}.
\end{align}
Here $I_{\text{f}}$ represents terms that include fermions.
Therefore only the lowest components in the K\"ahler tensor
$\Lambda_{ik\bar{j}\bar{l}} (\Phi, \Phi^{\dagger})$ contribute to the
purely bosonic parts of the Lagrangian.
Then the component Lagrangian is given by
\begin{align}
\mathcal{L} =& \
\frac{\partial^2 K}{\partial \varphi^i \partial \bar{\varphi}^{\bar{j}}}
\left(
- \partial_m \varphi^i \partial^m \bar{\varphi}^{\bar{j}} + F^i \bar{F}^{\bar{j}}
\right)
+ \frac{\partial W}{\partial \varphi^i} F^i
+ \frac{\partial\bar{W}}{\partial \bar{\varphi}^{\bar{j}}} \bar{F}^{\bar{j}}
\notag \\
 & \ +
\Lambda_{ik\bar{j} \bar{l}} (\varphi, \bar{\varphi})
\left\{
(\partial_m \varphi^i \partial^m \varphi^k) (\partial_n \bar{\varphi}^{\bar{j}}
\partial^n \bar{\varphi}^{\bar{l}}) - 2 \partial_m \varphi^i \partial^m
\bar{\varphi}^{\bar{j}} F^k \bar{F}^{\bar{l}} + F^i \bar{F}^{\bar{j}}
 F^k \bar{F}^{\bar{l}} \right\}
\notag \\
& \ + \mathcal{L}_{\text{fermions}},
\label{eq:off-shell_Lagrangian}
\end{align}
where $\mathcal{L}_{\text{fermions}}$ is terms that include fermionic fields.
Note that $\Lambda_{ik\bar{j}\bar{l}} (\Phi, \Phi^{\dagger})$
generically contains
space-time derivatives of the chiral superfields
and there are arbitrary order of derivative terms in the Lagrangian
\eqref{eq:off-shell_Lagrangian}.
The equation of motion for $\bar{F}$ is
\begin{align}
\frac{\partial^2 K}{\partial \varphi^i \partial \bar{\varphi}^{\bar{j}}}
 F^i - 2 \Lambda_{ik \bar{j} \bar{l}} \partial_m \varphi^i F^k
 \partial^m \bar{\varphi}^{\bar{l}}
+ 2 \Lambda_{ik \bar{j} \bar{l}} F^i F^k \bar{F}^{\bar{l}} +
 \frac{\partial \bar{W}}{\partial \bar{\varphi}^{\bar{j}}} = 0.
\label{eq:auxiliary_eom}
\end{align}
As we have advertised, this equation does not contain any space-time derivatives on $F$.
Then, in principle, the equation \eqref{eq:auxiliary_eom} is algebraically
solvable. However, it is not so straightforward to solve the equation for
general $N$ since it is a simultaneous equation of cubic order.
Only a few solutions have been known. For example, for $N=1$ single chiral superfield
models, one can solve the cubic order equation
\eqref{eq:auxiliary_eom} by the Cardano's method \cite{Sasaki:2012ka}.
Consequently, there are multiple distinct on-shell branches associated
with the independent solutions to the auxiliary fields.
To see this explicitly, let us begin with a single superfield model
without superpotential. The equation for $\bar{F}$ becomes
\begin{align}
K_{\varphi \bar{\varphi}} F
- 2  \Lambda \partial_m \varphi \partial^m \bar{\varphi} F + 2 \Lambda
 F^2 \bar{F} = 0,
\label{eq:auxiliary_eq_single}
\end{align}
where $K_{\varphi \bar{\varphi}}$ is the K\"ahler metric.
There are two kinds of solutions to this equation. One is the trivial solution
$F = 0$ and
the bosonic part of
the on-shell Lagrangian for this solution is
\begin{align}
\mathcal{L} =& \
- K_{\varphi \bar{\varphi}} \partial_m \varphi \partial^m \bar{\varphi}
+ \Lambda
(\partial_m \varphi \partial^m \varphi) (\partial_n \bar{\varphi}
 \partial^n \bar{\varphi}).
\label{eq:canonical}
\end{align}
We call the theory for the solution $F=0$ canonical branch.
The Lagrangian \eqref{eq:canonical} represents the ordinary quadratic kinetic term
of the complex scalar field $\varphi$ with the higher derivative interactions
governed by the tensor $\Lambda (\varphi,\bar{\varphi})$.
It is evident that the higher derivative corrections are introduced as a
perturbation to the quadratic kinetic term.

On the other hand, there is another non-trivial solution to the
auxiliary field equation \eqref{eq:auxiliary_eq_single}:
\begin{align}
F \bar{F} = - \frac{K_{\varphi \bar{\varphi}}}{2 \Lambda} + \partial_m
 \varphi \partial^m \bar{\varphi}.
\label{eq:non-canonical}
\end{align}
The bosonic part of the on-shell Lagrangian for the solution
\eqref{eq:non-canonical} is
\begin{align}
\mathcal{L} =
\left(
|\partial_m \varphi \partial^m \varphi|^2 - (\partial_m \varphi
 \partial^m \bar{\varphi})^2
\right) \Lambda - \frac{(K_{\varphi \bar{\varphi}})^2}{4 \Lambda}.
\label{eq:non-canonical}
\end{align}
In this branch, the quadratic canonical kinetic term disappears and the last term
is interpreted as a potential term. We call this non-canonical branch.
Compare to the canonical branch, the higher derivative terms are not
introduced perturbatively. We cannot take the limit $\Lambda \to 0$ in
this branch.

Even though the higher derivative interactions appear in a different way
in the Lagrangians \eqref{eq:canonical} and \eqref{eq:non-canonical}, SUSY is manifestly realized in each branch.
A specific feature, for example, BPS states in the single chiral
superfield models were discussed in Refs.~\cite{Adam:2011gc, Nitta:2014pwa
}.
The higher derivative corrections to the NG supermultiplets in
supersymmetric vacua were discussed in Ref.~\cite{Nitta:2014fca}.
For multiple or matrix-valued fields models, it is not so straightforward to solve
the equation for the auxiliary fields,
but only one example
 can be found in Ref.~\cite{Gudnason:2015ryh} in which
the authors solved the equation corresponding to Eq.~\eqref{eq:auxiliary_eom}
for the $SU(2)$-valued auxiliary field and found a supersymmetric extension of
the Skyrme model.

In the next section we focus on a single chiral superfield model and
discuss SUSY breaking in spatially modulated vacua.

\section{Spatially modulated vacuum in supersymmetric higher derivative model}\label{sec:modulation}
In this section, we investigate spatially modulated vacua in the
supersymmetric field theories with higher derivative terms.
For simplicity, 
we consider single superfield models without superpotential
where the K\"ahler metric is a constant $K_{\varphi
\bar{\varphi}} = k > 0$ and we focus on the canonical branch.
For a generic $\Lambda$, the energy density is given by
\begin{align}
\mathcal{E} =& \
 k (|\dot{\varphi}|^2 + |\partial_i
 \varphi|^2) + \Lambda 
\left\{
3 |\dot{\varphi}|^4 - \dot{\varphi}^2 (\partial_i \bar{\varphi})^2 -
 \dot{\bar{\varphi}}^2 (\partial_i \varphi)^2 - (\partial_i \varphi)^2
 (\partial_j \bar{\varphi})^2
\right\}
\notag \\
- & \
\frac{\partial \Lambda}{\partial \dot{\varphi}} |\dot{\varphi}|^2 
\left\{
(- \dot{\varphi}^2 + (\partial_i \varphi)^2) (- \dot{\bar{\varphi}}^2 +
 (\partial_i \bar{\varphi})^2 )
\right\}
\notag \\
- & \
\frac{\partial \Lambda}{\partial \dot{\bar{\varphi}}} |\dot{\varphi}|^2 
\left\{
(- \dot{\varphi}^2 + (\partial_i \varphi)^2) (- \dot{\bar{\varphi}}^2 +
 (\partial_i \bar{\varphi})^2 )
\right\},
\label{eq:energy}
\end{align}
where $i,j = 1,2,3$
and $\dot{\varphi} =
\frac{\partial\varphi}{\partial x^0}, \dot{\bar{\varphi}} =
\frac{\partial \bar{\varphi}}{\partial x^0}$.
Note that, in general, the Hermitian K\"ahler tensor $\Lambda$ is a
function of $\varphi, \partial_m \varphi$ and their Hermitian conjugate.
Vacua are defined such that the configurations minimize the energy
density $\mathcal{E}$.
We are interested in models where static, spatially modulated configurations are
realized as vacua.
Namely, we look for a K\"ahler tensor $\Lambda$ for which a spatial derivative of the field
$\varphi$ develops constant non-zero VEVs.
The simplest example is the one-dimensional spatial modulation.
In order to determine $\Lambda$ which realizes a modulated vacuum, we
assume the configuration $\dot{\varphi} = \partial_2 \varphi =
\partial_3 \varphi = 0$ and non-zero $\partial_1 \varphi$.
Then the energy density becomes
\begin{align}
\mathcal{E} = k |\partial_1 \varphi|^2 - \Lambda |\partial_1 \varphi|^4.
\label{eq:generic_energy_density}
\end{align}
Configurations $\partial_1 \varphi = \text{const.} \not= 0$ that minimize
\eqref{eq:generic_energy_density} are spatially modulated vacua along
the $x^1$-direction.
Since $\partial_1 \varphi$ appears as the absolute value in
\eqref{eq:generic_energy_density}, we further assume that $\Lambda$ is a
function of $|\partial_1 \varphi|$ only. 
This results in the situation where the shift symmetry $\varphi \to
\varphi + c$ is preserved. Here $c$ is a constant.
Then the energy density \eqref{eq:generic_energy_density},
which is a function of $X \equiv | \partial_1 \varphi| \ge 0$, is interpreted
as a potential for $X$:
\begin{align}
\mathcal{E} = k X^2 - \Lambda (X) X^4, \quad k > 0, X \ge 0.
\end{align}
One easily finds that for the simplest choice $\Lambda = \lambda = \text{const.}$, there are no
minima other than $X = 0$.
The next simplest choice of $\Lambda$ is 
\begin{align}
\Lambda = \lambda - \alpha |\partial_1 \varphi|^2,
\end{align}
where $\alpha$ is a real constant.
This corresponds to the choice
\begin{align}
\Lambda = \lambda - \alpha \partial_m \Phi \partial^m \Phi^{\dagger}.
\label{eq:proto}
\end{align}
Then the energy density for a one-dimensional modulation $\partial_1
\varphi \not= 0$ becomes
\begin{align}
\mathcal{E} =& \
\alpha X^6 - \lambda X^4 + k X^2.
\end{align}
As we will see below, for $\lambda > 0, \alpha>0$ there are
local (global) minima at $X \not= 0$.
Note that for this choice of $\Lambda$, 
the bosonic part of the Lagrangian \eqref{eq:canonical} becomes the one that
we studied in Ref.~\cite{Nitta:2017mgk} which allows for a spatially
modulated vacuum.
In the following, we make a brief
summary of the modulated vacuum found in Ref.~\cite{Nitta:2017mgk}.
We also note that although the theory manifestly realizes SUSY, the
energy functional \eqref{eq:energy} is not positive (semi) definite.
Therefore, vacua of the theory need not to have zero energy in general
even in supersymmetric theories.
Indeed, the spatially modulated vacuum allows the negative energy as we
will see below.

Since the energy density is a function of $|\partial_1 \varphi|^2$, it
is convenient to define $x \equiv |\partial_1 \varphi|^2$ and treat
$\mathcal{E}$ is a function of $x$:
\begin{align}
\mathcal{E} (x) \equiv \alpha x^3 - \lambda x^2 + k x, \qquad x \ge 0.
\end{align}
All minima of the function $\mathcal{E} (x)$ that satisfy the equation of
motion are vacua of the model.
At first, one finds the minimum $x=0$ in which the scalar field
has a constant or vanishing VEV.
In addition to this trivial vacuum, the function $\mathcal{E} (x) (x\ge
0)$ can have another minimum at $x \not= 0$ in which the space-time
derivative of $\varphi$ has non-zero VEVs.
This is indeed the case when the parameters $k, \lambda, \alpha$ satisfy the condition
$\lambda^2 - 3 \alpha k > 0$. The potential $\mathcal{E}(x)$ has a minimum at
\begin{align}
x_{+} =
\frac{\lambda + \sqrt{\lambda^2 - 3 \alpha k}}{3 \alpha},
\label{eq:xplus}
\end{align}
which is obviously non-zero.
At the vacuum $|\partial_1 \varphi|^2 = x_+$, we found the following
 spatially modulated configuration:
\begin{align}
\varphi (x^1) = \varphi_0 e^{i p x^1}, \qquad
\varphi_0,
 p \in \mathbb{R}, \
\label{eq:mod_vac}
\end{align}
where the constants $p, \varphi_0$ satisfy $p^2 \varphi_0^2 = x_{+}$.
The period of the modulation is given by $2 \pi/p$.
In the previous paper \cite{Nitta:2017mgk}, we have found that the
configuration \eqref{eq:mod_vac}
satisfies the equation of motion and it is a completely
consistent
vacuum of the theory.
The modulated vacuum
\eqref{eq:mod_vac} spontaneously breaks the translational symmetry along
the $x^1$-direction and the rotational symmetries in the $(x^1,x^2)$,
$(x^1,x^3)$ planes, as well as the $U(1)$ symmetry
 $\varphi \to e^{i \theta } \varphi$.
We have shown that there remain symmetries of the
$2+1$ dimensional Poincar\'e group $ISO(2,1)$ and a simultaneous
operations of the translation $P^1$ along the $x^1$-direction and the
$U(1)$ transformation $[P^1 \times U(1)]_{\text{sim}}$.
We have also pointed out that only the breaking pattern
$P^1 \times U(1) \to [P^1 \times U(1)]_{\text{sim}}$
gives rise to an NG boson.

\begin{figure}[t]
\begin{center}
\subfigure[$\mathcal{E} (x_+) > 0$ ($\alpha = 1, \lambda = 3.9, k = 4$)]
{
\includegraphics[scale=.57]{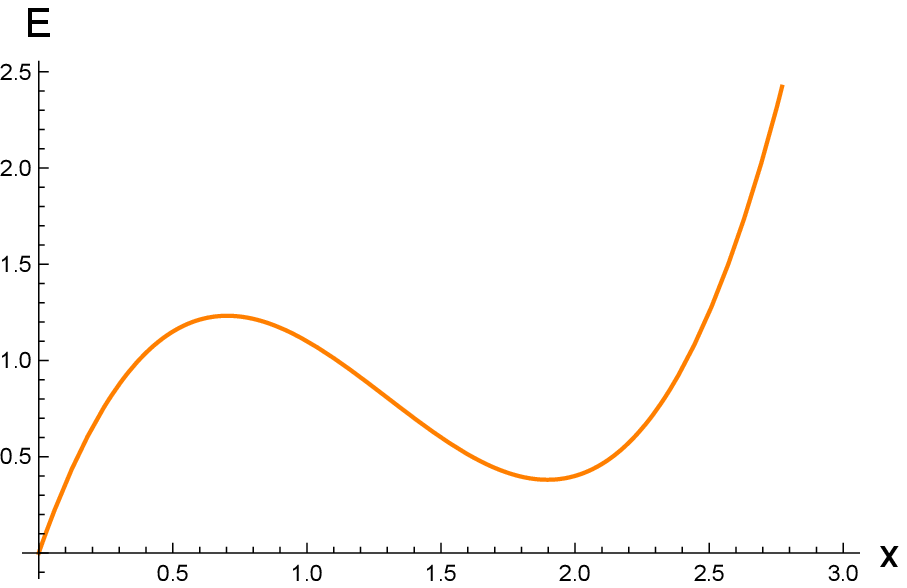}
}
\subfigure[$\mathcal{E} (x_+) = 0$ ($\alpha = 1,\lambda = 4, k = 4$)]
{
\includegraphics[scale=.57]{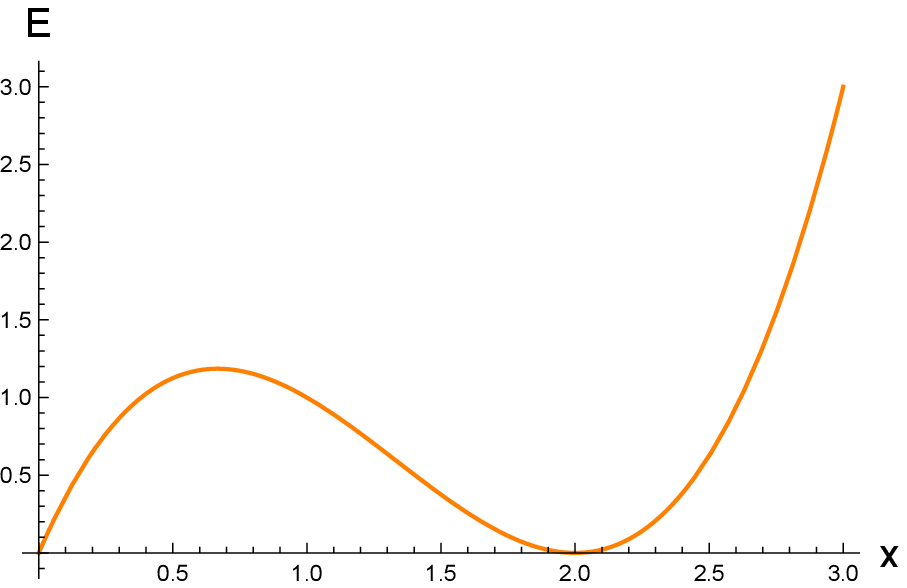}
}
\subfigure[$\mathcal{E} (x_+) < 0$ ($\alpha = 1, \lambda = 3, k = 2$)]
{
\includegraphics[scale=.57]{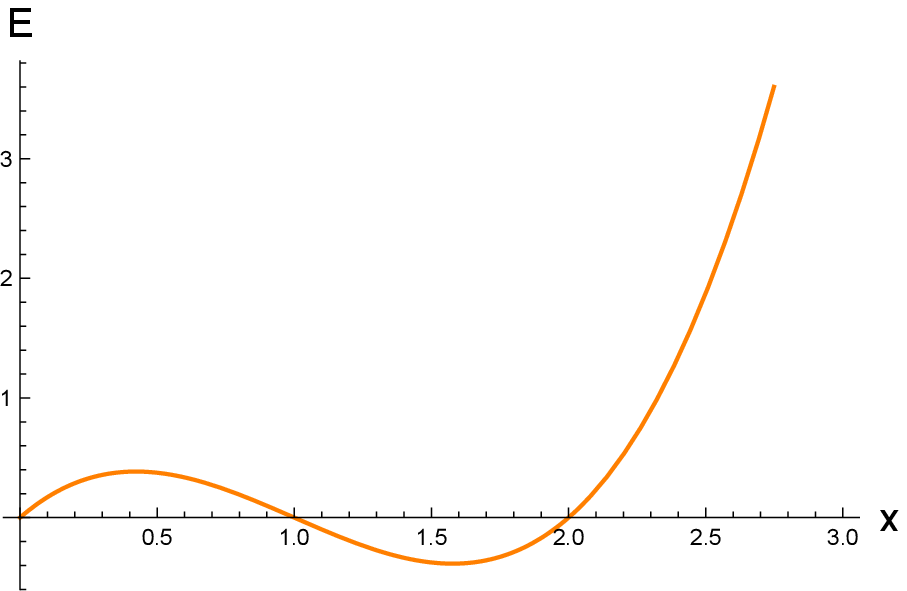}
}
\end{center}
\caption{Profiles of the energy function $\mathcal{E} (x)$.
The vertical and the horizontal axises stand for the energy $\mathcal{E}
 (x)$ and $x$.
The local vacua for (a) positive, (b) zero and (c) negative vacuum
 energies with examples of the parameters are shown.
}
\label{fig:energy_functional}
\end{figure}

In order to clarify the SUSY breaking in the modulated vacuum
\eqref{eq:mod_vac}, we recall the
SUSY variation of the fermion $\psi$.
In the canonical branch, this is given by
\begin{align}
\delta \psi_{\alpha} = \ i \sqrt{2} (\sigma^m)_{\alpha \dot{\alpha}}
 \bar{\xi}^{\dot{\alpha}} \partial_m \varphi + \sqrt{2} \xi_{\alpha} F
 (\varphi, \bar{\varphi})
= \ i \sqrt{2} \sigma^1 \bar{\xi} \partial_1 \varphi.
\label{eq:SUSY_breaking}
\end{align}
Here, $\xi, \bar{\xi}$ are parameters of the SUSY transformation.
It is clear that SUSY is preserved in the vacuum $x = 
|\partial_1 \varphi|^2 = 0$.
On the other hand, in the modulated vacuum \eqref{eq:mod_vac}, one finds that
the variation \eqref{eq:SUSY_breaking} does not vanish and SUSY is
spontaneously broken there.
A particular emphasis is placed on the fact that non-zero values of the auxiliary field $F$ is not an order
parameter of the SUSY breaking anymore.
This is a reflection of the fact that the energy density
\eqref{eq:energy} of this model is not positive (semi) definite.
In order to illustrate this issue, we examine the
sign of the vacuum energy $\mathcal{E}$.
The vacuum energy at $x = x_{+}$ is calculated as
\begin{align}
\mathcal{E} (x_+) = - \frac{1}{27 \alpha^2}
\left(
\lambda + \sqrt{\lambda^2 - 3 \alpha k}
\right)
\left\{
- 6 \alpha k + \lambda
\left(
\lambda + \sqrt{\lambda^2 - 3 \alpha k}
\right)
\right\}.
\label{eq:vac_energy}
\end{align}
It is evident that the quantity \eqref{eq:vac_energy} is not always positive semi-definite.
We have found that the sign of the vacuum energy is classified according
to the discriminant condition of the function $\alpha x^2 - \lambda x + k$.
Depending on the parameters $k,\lambda, \alpha$, we have three distinct types of vacua. In the
following, we assume that all the parameters satisfies the condition $\lambda^2 -
3 \alpha k > 0$ which guarantees that the potential has a local minimum
given in Eq.~\eqref{eq:xplus}.
\\
\\
\underline{\bf \textbullet \ Positive energy vacuum}
\\
\\
When the parameters satisfy the discriminant condition $\lambda^2 - 4
\alpha k < 0$, then the function $\mathcal{E} (x) = \alpha x^3 -
\lambda x^2 + k x
$ is
positive definite.
If this is the case, the potential function $\mathcal{E} (x)$ looks like
Fig.~\ref{fig:energy_functional} (a).
We find that the local vacuum energy at $x = x_+$ is positive
$\mathcal{E} (x_+) > 0$ and the SUSY breaking vacuum at $x = x_+$ is meta-stable.
It seems that the meta-stable vacuum decays to the global supersymmetric
vacuum at $x = 0$ within a finite time. However we can make the life-time of
the meta-stable vacuum longer by choosing parameters of the potential
appropriately.
If the life-time is longer than that of the Universe, this kind of
meta-stable vacuum becomes a possible candidate of
phenomenologically viable grand state.
Indeed, the dynamical SUSY breaking in a meta-stable vacua was
discussed in the framework of supersymmetric effective theories
\cite{Intriligator:2006dd}.
\\
\\
\underline{\bf \textbullet \ Zero energy vacuum}
\\
\\
When the parameters $k, \lambda, \alpha$ satisfy the condition
$\lambda^2 - 4 \alpha k = 0$,
then a schematic picture of the function $\mathcal{E}(x)$ is given by
Fig.~\ref{fig:energy_functional} (b).
In addition to the SUSY vacuum $x = 0$, we have a local vacuum $x = x_+$
in which $\mathcal{E} (x_+) = 0$. They are actually degenerated global
vacua.
Interestingly, although $ \mathcal{E} (x_+) = 0$, this does not imply that the vacuum
preserves SUSY. In fact,
we have seen that
SUSY is broken by the condition in Eq.~\eqref{eq:SUSY_breaking}.
This results in the fact that the Goldstino in this vacuum becomes
non-dynamical and does not propagate in the
directions  transverse to the
modulation as we will see later.
\\
\\
\underline{\bf \textbullet \ Negative energy vacuum}
\\
\\
Finally, we consider the condition $\lambda^2 - 4 \alpha k > 0$.
When this is the case, the function $\mathcal{E}(x)$ looks like
Fig.~\ref{fig:energy_functional} (c).
Now the supersymmetric vacuum $x = 0$ becomes meta-stable and the
SUSY breaking vacuum at $x = x_{+}$ is energetically favoured.
Therefore the
global
vacuum is located at $x = x_+$ in which $\mathcal{E}(x_+) < 0$.
In this vacuum, SUSY is again broken by the condition
\eqref{eq:SUSY_breaking}. We will discuss the Goldstone mode associated with the
SUSY breaking in the negative vacuum state in the next section.

\section{Nambu-Goldstone modes in supersymmetry breaking modulated vacuum}\label{sec:NG}
In this section, we study NG modes in the
SUSY-breaking spatially modulated vacuum \eqref{eq:mod_vac}.
There are two kinds of NG modes.
One is the bosonic mode which appears due to the
spontaneously broken symmetry $P^1 \times U(1) \to [P^1 \times U(1)]_{\text{sim}}$
in the modulated vacuum.
We note that the translation $P^1$ and the rotations in the $(x^1,x^2)$ and
$(x^1,x^3)$ planes are not independent each other \cite{Low:2001bw}.
This is a particular example of the inverse Higgs effect \cite{Ivanov:1975zq}. 
Therefore, there is only one bosonic NG mode.
The other is the fermionic NG mode (Goldstino)
associated with the SUSY breaking.
In the following, we discuss bosonic and fermionic NG modes
separately.

\subsection{Bosonic sector}
We first summarize the bosonic NG mode
in the modulated vacuum in the model characterized by \eqref{eq:proto},
which is identical to
the one studied in
Ref.~\cite{Nitta:2017mgk}.
We shift the field from the modulated vacuum \eqref{eq:mod_vac} and
introduce the fluctuation $\tilde{\varphi}$ as a dynamical field:
\begin{align}
\varphi \ \longrightarrow \ \langle \varphi \rangle + \tilde{\varphi},
\end{align}
where $\langle \varphi \rangle = \varphi_0 e^{i p x^1}$ is the
modulating VEV.
The quadratic terms of the dynamical scalar field $\tilde{\varphi}$ are
extracted from the Lagrangian \eqref{eq:canonical}.
The result is
\begin{align}
\mathcal{L}_{\text{quad.} \tilde{\varphi}} =
- \frac{1}{2} \vec{\varphi}^{\dagger} \mathbf{M} \vec{\varphi}.
\end{align}
Here we have defined the following quantities:
\begin{align}
\vec{\varphi} =
\left(
\begin{array}{c}
\partial_{\hat{m}} \tilde{\varphi} \\
\partial_{\hat{m}} \tilde{\varphi}^{\dagger} \\
\partial_1 \tilde{\varphi} \\
\partial_1 \tilde{\varphi}^{\dagger}
\end{array}
\right),
\qquad
\mathbf{M} =& \
\left(
\begin{array}{cc}
M_1 & \ 0 \\
0 & M_2
\end{array}
\right).
\end{align}
We have separated the terms to the $SO(2,1)$ invariant transverse sector
$(\hat{m} = 0,2,3)$ and the modulation sector.
In the $4 \times 4$
Hermitian
matrix $\mathbf{M}$, each block element is given by
\begin{align}
M_1 =& \
\left(
\begin{array}{cc}
k + \alpha x_{+}^2 & 2 (\lambda - \alpha x_{+}) x_+
 e^{- 2 i p x^1} \\
2 (\lambda - \alpha x_{+}) x_+ e^{2 i p x^1} & k + \alpha x_{+}^2
\end{array}
\right),
\notag \\
M_2 =& \
\left(
\begin{array}{cc}
9 \alpha x_+^2 - 4 \lambda x_+ + k & 2 (\lambda -3 \alpha x_{+}) x_+
 e^{- 2 i p x^1} \\
2 (\lambda - 3\alpha x_{+}) x_+ e^{2 i p x^1} & 9 \alpha x_+^2 - 4 \lambda x_+ + k
\end{array}
\right).
\end{align}
The eigenvalues of $M_1$ and $M_2$ determine the coefficients of the
quadratic kinetic terms in the $SO(2,1)$ invariant transverse and the
modulation directions, respectively. In our previous paper \cite{Nitta:2017mgk}, we
have found that $M_1$ and $M_2$ have zero and positive eigenvalues respectively.
The quadratic kinetic term for the zero eigenvalues modes vanish.
We pointed out that the mode associated with the zero eigenvalue of
$M_2$ in the modulation direction corresponds to the NG mode which
appears due to the spontaneous breaking of $P^1 \times U(1)$.
We have also shown that
cubic derivative terms for the bosonic NG mode are
absent and
a quartic derivative term of the NG mode
 appears in the Lagrangian.
On the other hand,
the positive eigenvalue mode in the $M_2$ sector is the Higgs mode
which has a quadratic kinetic term.
This is apparently a gapless mode.
This is a generalization of the ordinary NG theorem where the NG and the Higgs
modes appear as zero and positive eigenvalue modes for the quadratic
curvature of the potential energy. The difference
from the ordinary NG theorem
is that we have VEVs
for the derivative of fields but not fields
themselves.
The zero eigenvalue of $M_1$ in the $SO(2,1)$ invariant sector,
corresponds to a flat direction of the potential.

\subsection{Fermionic sector}
We next investigate fermions in the modulated vacuum.
The situation is quite different from the bosonic sector.
To see this, let us consider the $\mathcal{N} =1$ SUSY algebra:
\begin{align}
\{ Q_{\alpha}, \bar{Q}_{\dot{\alpha}} \} = 2 (\sigma^m)_{\alpha
 \dot{\alpha}} P_m,
\end{align}
where $Q_{\alpha}, \bar{Q}_{\dot{\alpha}}$ are supercharges and $P^m$ is
the generator of translation. Then, the energy for a state $|\Psi
\rangle$ is given by
\begin{align}
E_{\Psi} =
\langle \Psi | P^0 | \Psi \rangle =
\frac{1}{4} \sum_{\alpha, \dot{\alpha} = 1,2}
\left(
\| Q_{\alpha} |\Psi \rangle  \|^2
+
\| \bar{Q}_{\dot{\alpha}} |\Psi \rangle  \|^2
\right).
\label{eq:SUSY_energy}
\end{align}
From the expression \eqref{eq:SUSY_energy}, one finds that
when the energy for a state $|\Psi \rangle$ is negative $E_{\Psi} < 0$,
then there are negative norm states (ghosts) in the system.
In particular, for a vacuum $|0 \rangle$,
since
SUSY
is spontaneously broken there,
the states $Q_{\alpha} |0 \rangle, \bar{Q}_{\dot{\alpha}}|0 \rangle
\not= 0$ are identified with the Goldstino in the zero-momentum
associated with the SUSY breaking.
We therefore expect that there are ghost Goldstino in the negative
energy modulated vacuum.

\begin{figure}[t]
\begin{center}
\subfigure[\mbox{}]
{
\includegraphics[scale=.57]{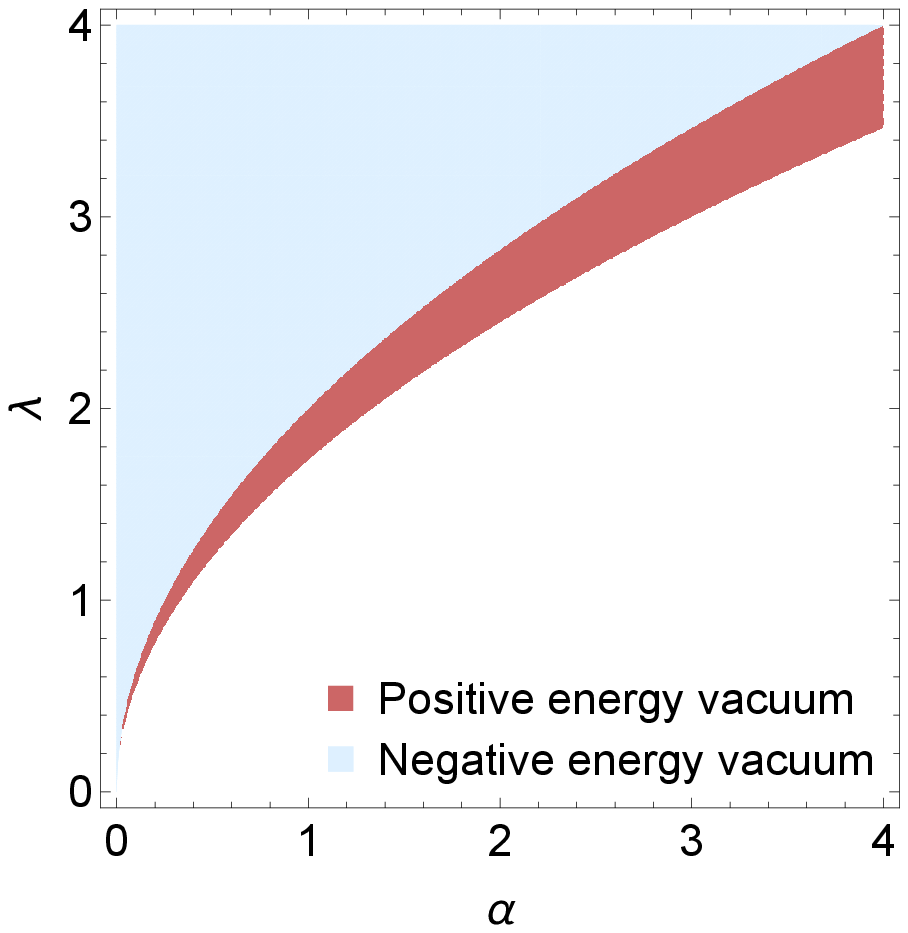}
}
\subfigure[\mbox{}]
{
\includegraphics[scale=.57]{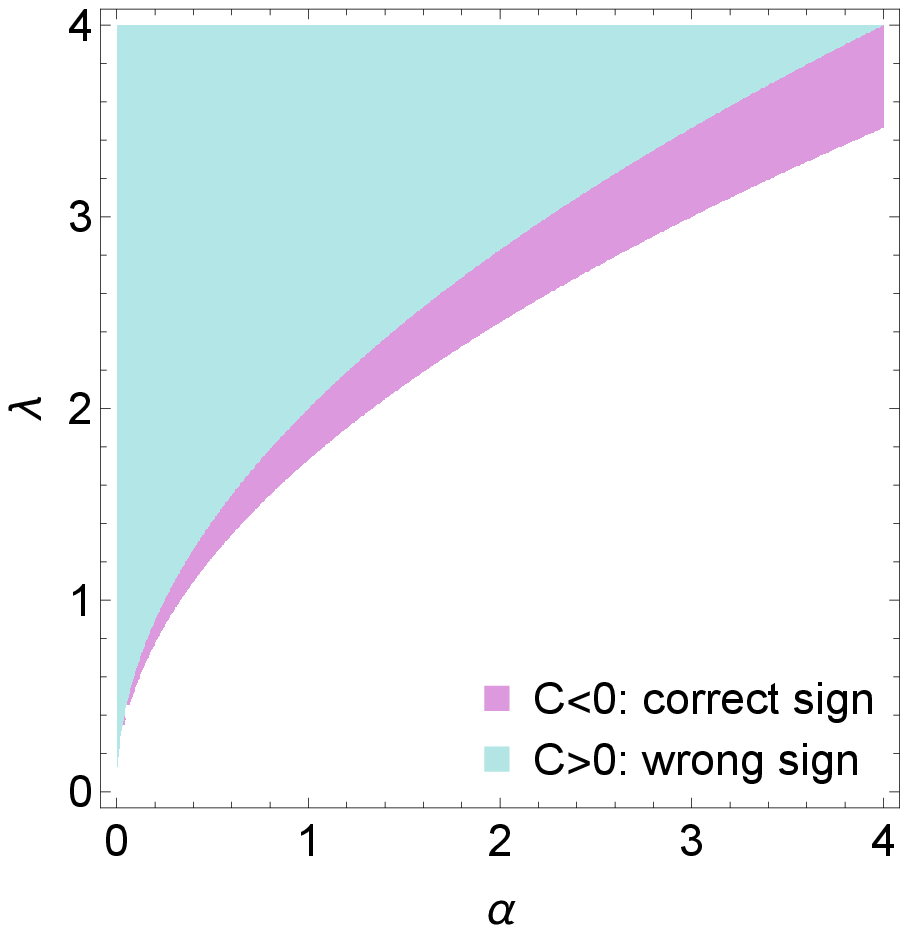}
}
\subfigure[\mbox{}]
{
\includegraphics[scale=.57]{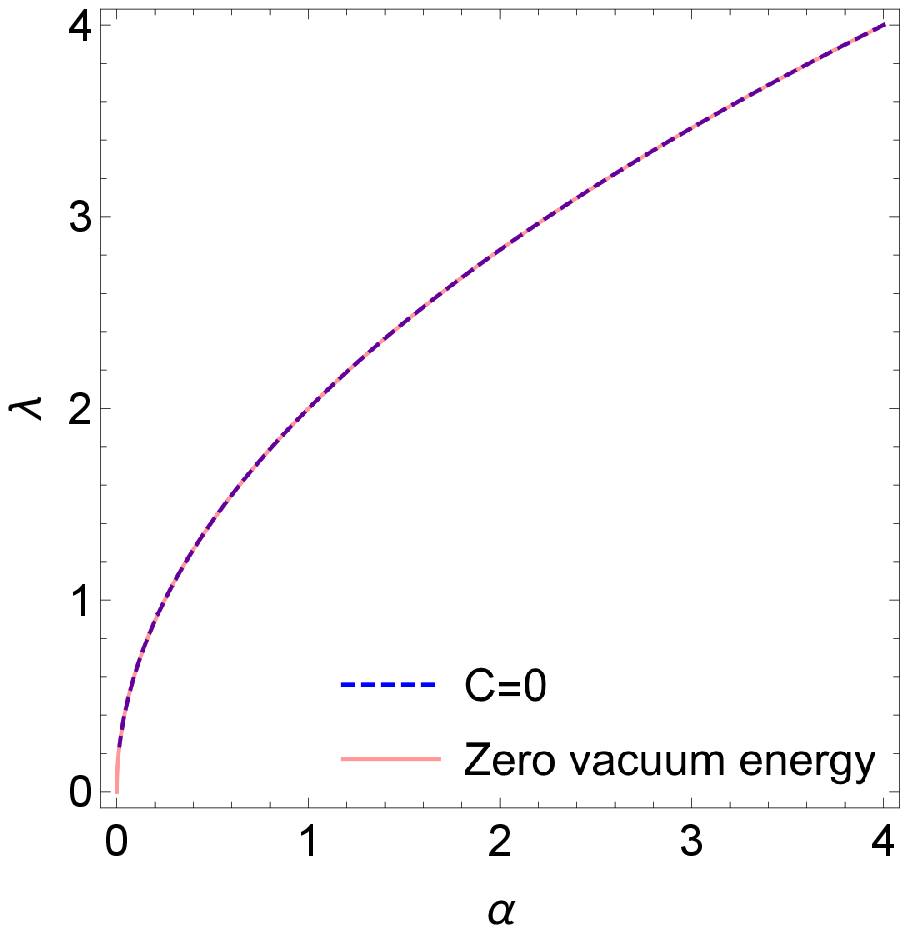}
}
\end{center}
\caption{
(a)
The parameter regions $(\alpha, \lambda)$ for the modulated vacuum with
 positive and negative energies.
(b)
The parameter region that the coefficient $C$ of the fermionic kinetic term
 has wrong sign $C>0$.
(c)
The parameter region (curves) of $(\alpha, \lambda)$ for the zero energy vacuum
  and the vanishing fermion kinetic term $C=0$.
Here all the examples are shown with $k = 1$ fixed.
}
\label{fig:vacua_regions}
\end{figure}

To see this explicitly, we evaluate the coefficient of the kinetic term
of $\psi$ in the chiral multiplet which is a unique candidate of the
Goldstino.
As one finds in Eq.~\eqref{eq:HD} in Appendix,
the fermion field in the Lagrangian appears with
the auxiliary field accompanied by the space-time derivative.
Eventually, the equation of motion for the auxiliary field becomes
non-algebraic when the fermion field is included.
In order to write down the quadratic kinetic term of the fermion in the
canonical branch, we solve the auxiliary field equation in the
perturbation of $\psi$.
Since the fermion emerges as a bi-linear in the solution of $F$,
we have $F = 0 + \mathcal{O} (\psi^2)$ in the canonical branch.
Using this fact,
the quadratic terms of the fermion $\psi$ in the Lagrangian are found
to be
\begin{align}
\mathcal{L}_{\text{quad.}\psi}
=& \
i \left\{
- k + (\lambda - \alpha x_+) x_+
\right\} \bar{\psi} \bar{\sigma}^{\hat{m}} \partial_{\hat{m}} \psi
\notag \\
&
+ \frac{i}{2}
\left\{
- k + 3 (\lambda - \alpha x_+) x_+ + 2 \alpha
x_{+}^2
e^{2 i p x^1}
\right\} \bar{\psi} \bar{\sigma}^1 \partial_1 \psi
\notag \\
&
+ \frac{i}{2}
 \left\{
- k + 3 (\lambda - \alpha x_+) x_+ + 2 \alpha
x_{+}^2
e^{- 2 i p x^1}
\right\} \psi \sigma^1 \partial_1 \bar{\psi}
\notag \\
&
+ p x_+
\left\{
\alpha x_+
-
(\lambda - \alpha x_+)
p x_+ \varphi_0
\right\}  \psi \sigma^1 \bar{\psi}.
\label{eq:fermion_kin}
\end{align}
Here
we have again separated the terms to the $SO(2,1)$ invariant transverse
and the modulation sectors.
The coefficient of the $SO(2,1)$ Lorentz invariant
fermion kinetic term
can be calculated as
\begin{align}
C
\equiv
- k + x_{+} (\lambda - \alpha x_{+}) =& \
 - k + \frac{1}{3 \alpha} (\lambda + \sqrt{\lambda^2 - 3 \alpha k})
 \left(
 \lambda - \frac{1}{3}
 ( \lambda +
 \sqrt{\lambda^2 - 3 \alpha k}
 \
 )
 \right).
\end{align}
Whether the fermion becomes a ghost or not can be read off from the
sign of the coefficient $C$.
When $C > 0$ ($C<0$), this is the wrong (correct) sign of
fermionic kinetic term,
and then $\psi$ is (not) a fermionic ghost.
The sign of $C$ is determined by the parameters $k,\lambda, \alpha$ which
are related to the sign of the vacuum energy we have classified before.
The parameter regions of the positive and negative vacuum energies
are found in Fig.~\ref{fig:vacua_regions} (a).
The regions that the coefficient of the fermion kinetic term $C$ has the correct $C < 0$ and
the wrong $C > 0$ signs are shown in Fig.~\ref{fig:vacua_regions} (b).
One finds that the regions of the positive energy and the correct sign
$C<0$ and those of the negative energy and the wrong sign
$C > 0$ completely coincide.
With this result at hand, we find that the Goldstino propagates in the
transverse direction in the correct way, {\it i.e.} it never becomes a
ghost in the meta-stable modulated vacuum.
On the other hand, the Goldstino becomes a ghost in the negative energy vacuum.
This is consistent with the SUSY algebra in Eq.~\eqref{eq:SUSY_energy}.
The norm of the Goldstino is positive (negative)
for positive (negative) vacuum energies.

Since the sign of $C$ changes continuously, one
notices that at the boundary of two regions, the kinetic term vanishes.
Indeed, the parameter curves for the zero vacuum energy and $C = 0$
coincides as in Fig.~\ref{fig:vacua_regions} (c).
We therefore expect that the Goldstino becomes non-dynamical
in the zero energy vacuum.
This is a conceivable result from the observation of the SUSY algebra
in Eq.~\eqref{eq:SUSY_energy}.
The fact that SUSY is broken in the zero vacuum energy results in the
relation
\begin{align}
0 = \sum_{\alpha, \dot{\alpha} = 1,2}
\left(
\| Q_{\alpha} |0 \rangle  \|^2
+
\| \bar{Q}_{\dot{\alpha}} |0 \rangle  \|^2
\right),
\end{align}
for the state $Q_{\alpha} |0 \rangle \not= 0$. Namely, the Goldstino
becomes a zero norm state and it disappears from the physical sector.
This is quite different from the ordinary SUSY breaking.

Things get more involved when we look at the kinetic term in the
modulation direction.
In order to clarify the sign of the coefficient of the kinetic term in the
modulation direction, we perform the partial integration in the third term in
Eq.~\eqref{eq:fermion_kin}.
We then find
\begin{align}
\mathcal{L}_{\text{quad.} \psi} =& \
i C \bar{\psi} \bar{\sigma}^{\hat{m}} \partial_{\hat{m}} \psi
+
i C_{\text{mod}} \bar{\psi} \bar{\sigma}^1 \partial_{1} \psi
\notag \\
&
+ p x_+^2
\left\{
\alpha
-  (\lambda - \alpha x_+) p \varphi_0
- 2 p e^{-2ipx^1}
\right\} \psi \sigma^1 \bar{\psi}.
\end{align}
Here we have defined
\begin{align}
C_{\text{mod}} \equiv
C - 2 \alpha x_+^2 (1 - \cos (2 p x^1)).
\end{align}
Due to the modulated vacuum, the coefficient $C_{\text{mod}}$ oscillates
in the $x^1$-direction.
However, since the inequality $C_{\text{mod}} \le C$ always holds, the
coefficient $C_{\text{mod}}$ takes negative values
in the parameter region for $C < 0$.
We therefore conclude that the Goldstino in the positive energy
vacuum propagates in the correct way even in the modulation
direction. Then the modulated vacuum with positive energy we found is
completely consistent (meta-)stable vacuum even in the fermionic sector.

On the other hand, because the minimum value of $C_{\text{mod}}$
\begin{align}
\min
C_{\text{mod}}
 = - k + (\lambda - \alpha x_+) x_+ - 4 \alpha x_+^2
= - 2 \alpha x_+^2 - \lambda x_+ < 0,
\end{align}
is negative even in the region for $C \ge 0$,
the modulation direction can have correct sign of the fermionic kinetic
term even in the negative or zero energy vacua.
This also indicates the fact that the Goldstino has non-zero kinetic
term along the modulated direction even in the zero energy vacuum.
Presumably, this is because the modulated vacuum \eqref{eq:mod_vac}
breaks the translational symmetry along $x^1$.
We can perform the Lorentz boost of the zero-momentum Goldstino
$Q_{\alpha} |0 \rangle$, whatever it is a ghost or not,
to obtain the one that has a non-zero momentum $P^{\hat{m}}$.
The resulting Goldstino has non-zero kinetic term $\bar{\psi} \bar{\sigma}^{\hat{m}}
\partial_{\hat{m}} \psi$ in the $SO(2,1)$ Lorentz invariant sector.
Since the sign of the norm does not change under the Lorentz
transformation, there is a one-to-one correspondence between
the sign of $C$ and the norm of $Q_{\alpha} |0 \rangle$ in the Lorentz invariant sector.
However, this discussion does not hold in the modulated direction.
We are not able to perform the translational transformation along the
$x^1$-direction to obtain $\psi$ that has a non-zero kinetic
term $\bar{\psi} \bar{\sigma}^{1}\partial_{1} \psi$.
Therefore the sign of $C_{\text{mod}}$ does not help in judging the existence of ghosts in
the modulated direction. 
In summary, we cannot say anything about ghosts in the modulated direction.

\section{Analysis with superpotential}\label{sec:superpot}
In this section we introduce an example of the
higher derivative chiral
model where a superpotential $W$ exists.
We demonstrate that superpotentials generically change the ``potential'' of
the derivative terms and a variety of
modulated vacua is possible.
The equation of motion for the auxiliary field in the single superfield
model
 with general $\Lambda$
becomes
\begin{align}
& K_{\varphi \bar{\varphi}} F + \left(
- 2 F \partial_m \varphi \partial^m \bar{\varphi} + 2 F^2 \bar{F}
\right) \Lambda + \frac{\partial \bar{W}}{\partial \bar{\varphi}} = 0,
\end{align}
where we have introduced only the bosonic fields.
After eliminating $\bar{F}$
in the above equation,
we have the equation only for $F$:
\begin{align}
2 \Lambda \frac{\partial W}{\partial \varphi} F^3 + \frac{\partial
 \bar{W}}{\partial \bar{\varphi}} (K_{\varphi \bar{\varphi}} - 2 \Lambda \partial_m \varphi
 \partial^m \bar{\varphi}) F + \left( \frac{\partial \bar{W}}{\partial
 \bar{\varphi}} \right)^2 = 0.
\end{align}
The solutions are given by the Cardano's formula:
\begin{align}
& F^{(a)} = \omega^a
\sqrt[3]{
- \frac{q}{2} +
\sqrt{
\left(
\frac{q}{2}
\right)^2 +
\left(
\frac{p}{3}
\right)^3
}
}
+ \omega^{3-a}
\sqrt[3]{
- \frac{q}{2}
- \sqrt{
\left(
\frac{q}{2}
\right)^2 +
\left(
\frac{p}{3}
\right)^3
}
}.
\label{eq:superpot_sol}
\end{align}
Here $\omega^3 = 1$ and $a = 0,1,2$. We have defined the following
quantities:
\begin{align}
p =& \ \frac{1}{2 \Lambda}
\left(
\frac{\partial W}{\partial \varphi}
\right)^{-1}
\left(
\frac{\partial \bar{W}}{\partial \bar{\varphi}}
\right)
\left(
K_{\varphi \bar{\varphi}} - 2 \Lambda \partial_m \varphi \partial^m \bar{\varphi}
\right), \\
q =& \ \frac{1}{2 \Lambda}
\left(
\frac{\partial W}{\partial \varphi}
\right)^{-1}
\left(
\frac{\partial \bar{W}}{\partial \bar{\varphi}}
\right)^2.
\end{align}
The purely bosonic terms of the on-shell Lagrangian is calculated as
\begin{align}
\mathcal{L}^{(a)} =& \
- K_{\varphi \bar{\varphi}} \partial_m \varphi \partial^m \bar{\varphi} + \Lambda (\partial_m
 \varphi \partial^m \varphi) (\partial_n \bar{\varphi} \partial^n
 \bar{\varphi})
\notag \\
& \
+ F^{(a)} \bar{F}^{(a)}
\left(
- K_{\varphi \bar{\varphi}} + 2 \Lambda \partial_m \varphi \partial^m \bar{\varphi}
\right) - 3 (F^{(a)} \bar{F}^{(a)})^2 \Lambda.
\end{align}
Here $F^{(a)},\bar{F}^{(a)}$ are one of the solutions for $a=0,1,2$ in
Eq.~\eqref{eq:superpot_sol}.
Apparently there are three distinct branches corresponding to $a=0,1,2$.

\begin{figure}[tb]
\begin{center}
\includegraphics[scale=0.55]{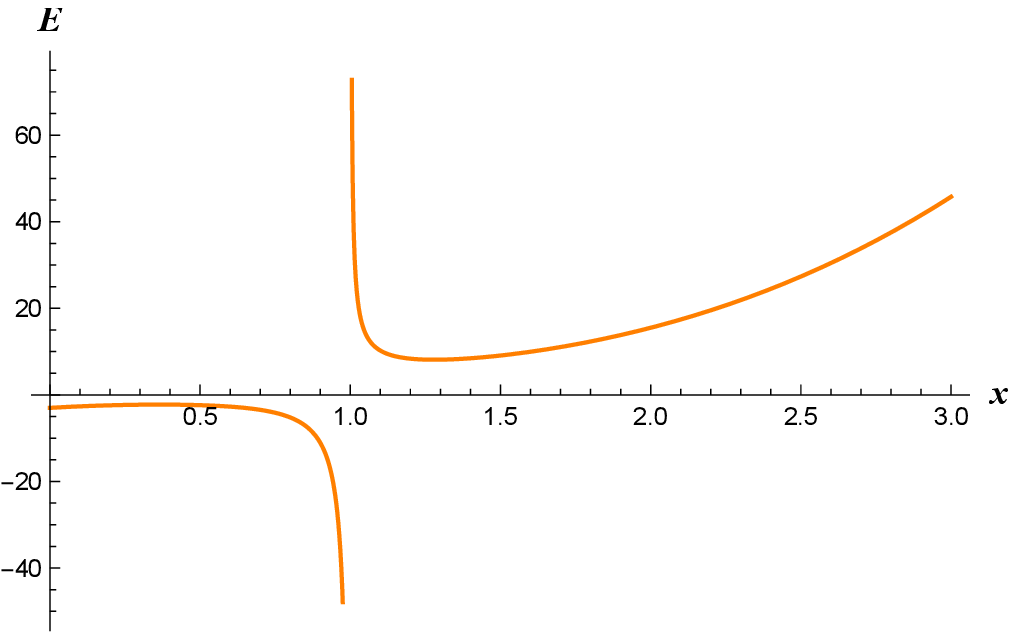}
\includegraphics[scale=0.55]{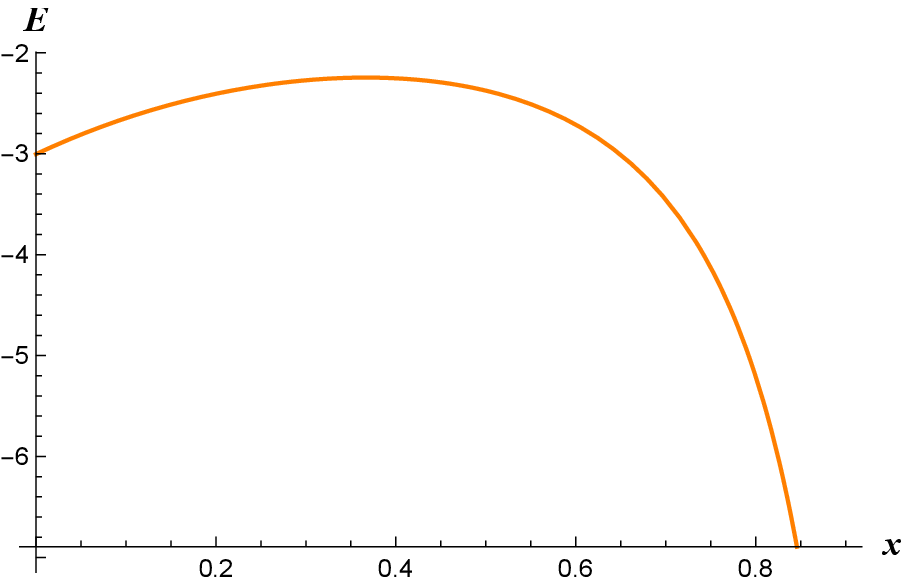}
\includegraphics[scale=0.55]{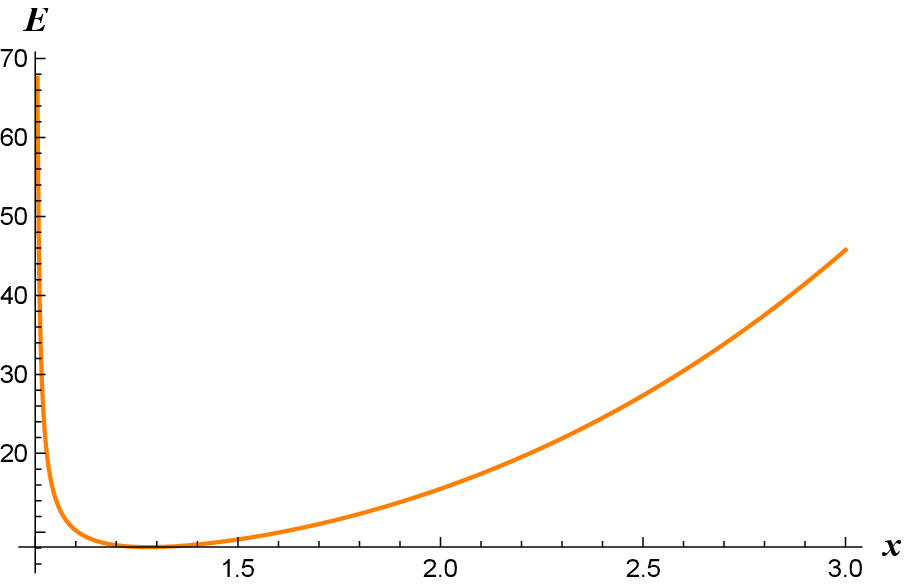}
\end{center}
\caption{
Energy plot for $\Lambda = \lambda - \alpha |\partial_m
 \varphi|^2, W =\beta \Phi$ with $k=1,
 \lambda = 1, \alpha = 1, \beta = 1$.
Left : The global structure of the energy functional $\mathcal{E}$. 
Middle : Enlarged view of $\mathcal{E}$ around the origin $0 \le x
 \le 0.9$. Right : Enlarged view $\mathcal{E}$ in the region $x \ge 1$.
The vertical and the horizontal axes represent the energy $\mathcal{E}
 (x)$ and $x = |\partial_1 \varphi|^2$.}
\label{fig:codim1W2}
\end{figure}

For simplicity, we choose the $a=0$ branch and employ the ansatz for static,
one-dimensional spatial configurations along $x^1$-direction, $\varphi =
\varphi (x^1)$. Then the energy functional becomes
\begin{align}
\mathcal{E}
=& \ K_{\varphi \bar{\varphi}} |\partial_1 \varphi|^2 - \Lambda
 (\partial_1 \varphi)^2 (\partial_1 \bar{\varphi})^2 - |F^{(0)}|^2
(K_{\varphi \bar{\varphi}} - 2 \Lambda |\partial_1 \varphi|^2) - 3 |F^{(0)}|^4 \Lambda.
\end{align}
Again, we consider the model
characterized by the tensor \eqref{eq:proto}
with the following simplest superpotential:
\begin{align}
W = \beta \Phi.
\end{align}
Here $\beta$ is a real constant.
The energy functional becomes a function of $x = |\partial_1 \varphi|^2$:
\begin{align}
\mathcal{E} = k x - (\lambda 
- \alpha x) x^2 - |F(x)|^2
 (k - 2 (\lambda - \alpha x) x) - 3 (\lambda - \alpha x) |F(x)|^4.
\end{align}
The auxiliary field in the $a=0$ branch is
\begin{align}
F(x) =& \
\left[
- \frac{\beta}{4} (\lambda - \alpha x)^{-1} +
\sqrt{
\frac{\beta^2}{16} (\lambda - \alpha x)^{-2}
+ \frac{1}{6^3} (\lambda - \alpha x)^{-3} (k - 2 (\lambda - \alpha x)x)^3
}
\right]^{\frac{1}{3}}
\notag \\
& \
+
\left[
- \frac{\beta}{4} (\lambda 
- \alpha x)^{-1} -
\sqrt{
\frac{\beta^2}{16} (\lambda -
 \alpha x)^{-2}
+ \frac{1}{6^3} (\lambda - \alpha x)^{-3} (k - 2 (\lambda - \alpha x)x)^3
}
\right]^{\frac{1}{3}}.
\end{align}
A schematic picture of $\mathcal{E}(x)$ is found in Fig \ref{fig:codim1W2}.
One finds that for the region in $0 < x < 1$, the potential is not bounded
from below and the system becomes unstable
(see the middle figure in Fig.~\ref{fig:codim1W2}).
The origin is a meta-stable supersymmetric vacuum (although the vacuum
energy in the example in Fig.~\ref{fig:codim1W2} is negative and we
expect that a ghost appears there).
However, in the region $x > 1$,  there is a global vacuum around 
$x = |\partial_1 \varphi|^2 \sim 1.280$ where SUSY is spontaneously broken
(see the right figure in Fig.~\ref{fig:codim1W2}).
The vacuum at $x \sim 1.280$ is clearly stable against decay and it has
positive energy ($\mathcal{E} \sim 8.126$).
This is an acceptable, stable supersymmetry breaking vacuum.
Even though the system becomes unstable in the small
$x$ region, this is an example where superpotential drastically changes
the stability of modulated vacua.

\section{Summary and discussions}\label{sec:summary}
In this paper, we have studied the spatially modulated vacua in
a supersymmetric theory with higher derivative terms.
We have focused on the model where the famous Ostrogradsky instability is absent.
Even though the scalar fields in the chiral multiplet appear with higher
derivatives, the model exhibits no propagating auxiliary fields.
The higher derivative part of the theory is defined by the
K\"ahler tensor $\Lambda$.
There are distinct on-shell branches corresponding to the different
solutions to the equation of motion for the auxiliary field.
We first consider the canonical branch in the model where the K\"ahler tensor
$\Lambda$ is given in Eq.~\eqref{eq:proto} and no superpotential.
The energy functional of this model is determined by the derivative
terms of the scalar fields.
We have found that the potential for the derivative terms allows
a local vacuum where SUSY is spontaneously broken.
In the SUSY breaking vacuum, we have shown that the
translational symmetry along one direction and
the rotational symmetries in the $(x^1,x^2), (x^1,x^3)$ planes
are broken. However the simultaneous transformation of $P^1$ and $U(1)$ is preserved in the
modulated vacuum. This modulated vacuum is completely consistent with
the equation of motion.
The vacuum energy depends on the parameters of the
K\"ahler
 metric and
tensor. There are vacua where the vacuum energy is positive, zero and
negative.

We have demonstrated that the quadratic canonical kinetic term for the
bosonic NG mode associated with the breaking of $P^1 \times
U(1)$ vanishes while
the Higgs boson
that are orthogonal
to the NG mode remains non-zero with correct sign.
This is a generalization of the NG theorem in higher
derivative theories.
On the other hand, the nature of the NG fermion
(Goldstino) in the modulated vacua is quite different from the bosonic modes.
We have found the SUSY breaking vacua where the vacuum energies
take
 positive, negative and zero values.
For the positive vacuum energy, the modulated vacuum is meta-stable
against decaying to the global supersymmetric vacuum.
However, sufficiently large possibilities of allowed parameters
$k,\alpha, \Lambda$ for the meta-stable vacuum indicate that one can
make the decay rate be so small compared with the life time of the
Universe \cite{Intriligator:2006dd}.
The Goldstino in this vacuum is well-behaved, namely, it has correct
sign of the kinetic term both in the $SO(2,1)$ Lorentz invariant and the modulated sectors.
We have also shown that when SUSY is spontaneously broken in the
vacuum where the vacuum energy is zero, then the kinetic term of the
Goldstino vanishes
and it becomes non-dynamical.
This is consistent with the SUSY algebra
in which the norm of the zero-momentum Goldstino states becomes zero.
This is quite different from the ordinary supersymmetric theories where
the zero energy vacuum corresponds to supersymmetric vacuum.
For the negative vacuum energy, the modulated vacuum is the global
minimum and it is the true vacuum.
The SUSY algebra together with the negative vacuum energy
implies that the Goldstino has a negative norm, {\it i.e.} it becomes a
fermionic ghost. We have explicitly shown that there appears the wrong sign for
the kinetic term of $\psi$ in the negative energy vacuum.
Although, goldstinos accompanied by the negative vacuum energy
are problematic in a physical theory \cite{Kugo:2017qma}
, there are several ways to remove undesirable ghost states from the physical
sector \cite{Ohta:1981bi,Ohta:1982ys, Fujimori:2016udq}.
We therefore conclude that the spatially modulated state with
positive vacuum energy is the physically acceptable supersymmetry
breaking vacuum in our model. 

We have also studied a model with a superpotential.
Although the on-shell Lagrangian is complicated due to the solution to
the equation of motion for the auxiliary field, we
have been able to
explicitly draw the potential energy for the derivative terms.
As an example, a simple model where the linear superpotential is introduced to the
prototypical model is analyzed. We
have found
that at large $x = |\partial_1
\varphi|^2$, there is a modulated vacuum which is stable against decaying.
We expect there are no ghost Goldstino in this vacuum.
However in the vicinity of the origin, the energy is not bounded from
below and the system suffers from the serious instability and ghosts.
In particular, in the supersymmetric vacuum in the origin, we expect
a ghost Goldstino. 
Alternative choices of $K, \Lambda$ and $W$ would help us to find a 
modulated vacuum which is the global minimum and has positive energy.

We have explicitly shown that the spontaneous SUSY breaking on a
spatially inhomogeneous vacuum actually occurs in a simple SUSY model where no propagating auxiliary
fields and no Ostrogradsky's ghost \cite{Ostrogradski} exist.
We stress that the spontaneous SUSY breaking in the spatially modulated
vacua -- that attract the greater attentions recently
\cite{Nakano:2004cd,Nickel:2009ke,Buballa:2014tba} --
, together with the ubiquity of the Lorentz violation \cite{Kostelecky:2003fs},
opens up robust possibilities of model buildings for particle physics
and cosmology \cite{Nojiri:2010wj, Cheng:2006us}.

Before closing the paper, we give several discussions.
In this paper we have 
discussed a new mechanism for spontaneous supersymmetry breaking based
on the modulated vacua studied in \cite{Nitta:2017mgk}. 
There are several interesting issues on the modulated vacua in
supersymmetric theories.
In this paper we have studied spatially modulated vacua only along one
direction. However, it is possible to find higher dimensional
modulation \cite{Kojo:2011cn}. It is also interesting to find
a temporal modulation \cite{Hayata:2013sea}.
It is conceivable that modulated vacua including the one presented in
this paper are ubiquitous in supersymmetric higher derivative theories.
We expect that these kinds of modulated vacua are utilized for phenomenological model buildings.
Embedding to supergravity \cite{Khoury:2010gb, Farakos:2012qu} is one of
the future directions.

Most notably, it is always true that the ordinary
quadratic kinetic term of the bosonic NG modes disappear in the modulated
vacua and there are derivative interactions of quartic type
\cite{Nitta:2017mgk}.
Although, these quartic derivative interactions do not show any
problematic behavior in the classical mechanics, they may cause some
(yet unknown) problems in quantum regime.
To our knowledge, there are no systematic analysis on consistent quantum
theories for such a vanishing quadratic kinetic term model.
It would be therefore interesting to study a quantum mechanical model where no
quadratic kinetic term of dynamical variables exist.
We will come back to these issues in future researches.

\subsection*{Acknowledgments}

The work of M.~N.~is
supported in part by
the Ministry of Education,
Culture, Sports, Science (MEXT)-Supported Program for the Strategic
Research Foundation at Private Universities `Topological Science' (Grant No.\ S1511006),
the Japan Society for the Promotion of Science
(JSPS) Grant-in-Aid for Scientific Research (KAKENHI Grant
No.~16H03984), and a Grant-in-Aid for Scientific Research on Innovative
Areas ``Topological Materials Science'' (KAKENHI Grant No.~15H05855) from the MEXT of Japan.
The work of S.~S. is supported in part by JSPS KAKENHI Grant Number
JP17K14294 and Kitasato University Research Grant for Young Researchers.
The work of R.~Y.~is supported by Research Fellowships of JSPS
 for Young Scientists Grant Number 16J03226.

\begin{appendix}

\section{Component expansion of the higher derivative terms}
The component expansion including fermions of the $\mathcal{N}=1$ chiral superfield in the
central basis is
\begin{align}
\Phi =& \ \varphi + i (\theta \sigma^m \bar{\theta}) \partial_m \varphi +
 \frac{1}{4} \theta^2 \bar{\theta}^2 \Box \varphi
 + \sqrt{2} \theta^{\alpha} \psi_{\alpha} - \frac{i}{\sqrt{2}}
 \theta^2 \partial_m \psi^{\alpha} (\sigma^m)_{\alpha \dot{\alpha}}
 \bar{\theta}^{\dot{\alpha}} + \theta^2 F
\notag \\
\Phi^{\dagger} =& \ \bar{\varphi} - i (\theta \sigma^m \bar{\theta})
 \partial_m \bar{\varphi} + \frac{1}{4} \theta^2 \bar{\theta}^2 \Box
 \bar{\varphi}
 + \sqrt{2} \bar{\theta}_{\dot{\alpha}} \bar{\psi}^{\dot{\alpha}} +
 \frac{i}{\sqrt{2}} \bar{\theta}^2 \theta^{\alpha} (\sigma^m)_{\alpha
 \dot{\alpha}} \partial_m \bar{\psi}^{\dot{\alpha}} + \bar{\theta}^2 \bar{F}.
\end{align}
The component expansion of the higher derivative term is \cite{Khoury:2010gb}:
\begin{align}
\frac{1}{16} (D\Phi)^2 (\bar{D} \Phi^{\dagger})^2
=& \
\theta^2 \bar{\theta}^2
\left[ \frac{}{} \right.
(\partial_m \varphi)^2 (\partial_n \bar{\varphi})^2 - 2 \bar{F} F
 \partial_m \varphi \partial^m \bar{\varphi} + \bar{F}^2 F^2
\notag \\
& \ - \frac{i}{2} (\psi \sigma^m \bar{\sigma}^n \sigma^p \partial_p
 \bar{\psi}) \partial_m \varphi \partial_n \bar{\varphi}
+ \frac{i}{2} (\partial_p \psi \sigma^p \bar{\sigma}^m \sigma^n
 \bar{\psi}) \partial_m \varphi \partial_n \bar{\varphi}
\notag \\
& \ + i \psi \sigma^m \partial^n \bar{\psi} \partial_m \varphi
 \partial_n \bar{\varphi} - i \partial^m \sigma^n \bar{\psi} \partial_m
 \varphi \partial_n \bar{\varphi} + \frac{i}{2} \psi \sigma^m \bar{\psi}
\left(
\partial_m \bar{\varphi} \Box \varphi - \partial_m \varphi \Box \bar{\varphi}
\right)
\notag \\
& \ + \frac{1}{2}
\left(
F \Box \varphi - \partial_m F \partial^m \varphi
\right) \bar{\psi}^2 + \frac{1}{2} (\bar{F} \Box \bar{\varphi} -
 \partial_m \bar{F} \partial^m \bar{\varphi}) \psi^2
\notag \\
& \ + \frac{1}{2} F \partial_m \varphi (\bar{\psi} \bar{\sigma}^m
 \sigma^n \partial_n \bar{\psi} - \partial_n \bar{\psi} \bar{\sigma}^m
 \sigma^n \bar{\psi}) + \frac{1}{2} \bar{F} \partial_m \bar{\varphi}
\left(
\partial_n \sigma^n \bar{\sigma}^m \psi - \psi \sigma^n \bar{\sigma}^m
 \partial_n \psi
\right)
\notag \\
& \ + \frac{3}{2} i \bar{F} F (\partial_m \psi \sigma^m \bar{\psi} - \psi
 \sigma^m \partial_m \bar{\psi})
+ \frac{i}{2} \psi \sigma^m \bar{\psi} (F \partial_m \bar{F} - \bar{F}
 \partial_m F)
\left. \frac{}{} \right]
\notag \\
& \ + \sqrt{2} i \bar{\theta}^2 (\partial_m \varphi)^2 (\theta \sigma^n
 \bar{\psi}) \partial_n \bar{\varphi}
- \sqrt{2} i \theta^2 (\partial_m \bar{\varphi})^2 (\psi \sigma^n
 \bar{\theta}) \partial_n \varphi
\notag \\
& \ + \sqrt{2} \theta^2 F \partial_m \bar{\varphi}
\left(
i \bar{F} (\psi \sigma^m \bar{\theta})
+
(\bar{\theta} \bar{\sigma}^m \sigma^n \bar{\psi}) \partial_m \varphi
\right)
\notag \\
& \
+
\sqrt{2} \bar{\theta}^2 \bar{F} \partial_m \varphi
\left(
- i F (\theta \sigma^m \bar{\psi})
+ (\psi \sigma^m \bar{\sigma}^n \theta) \partial_n \bar{\varphi}
\right)
\notag \\
& \ - \frac{1}{2} \bar{\theta}^2 (\partial_m \varphi)^2 \bar{\psi}
 \bar{\psi} - \frac{1}{2} \theta^2 (\partial_m \bar{\varphi})^2 \psi
 \psi + 2 (\psi \sigma^m \bar{\theta}) (\theta \sigma^n \bar{\psi})
 \partial_m \varphi \partial_n \bar{\varphi}
\notag \\
& \ + 2 \bar{F} F (\theta \psi) (\bar{\theta} \bar{\psi}) + i (\theta
 \sigma^m \bar{\theta})
(
F \partial_m \varphi \bar{\psi} \bar{\psi} - \bar{F} \partial_m
 \bar{\varphi} \psi \psi
)
+ \frac{1}{2} \theta^2 F^2 \bar{\psi} \bar{\psi} + \frac{1}{2}
 \bar{\theta}^2 \bar{F}^2 \psi \psi
\notag \\
& \ + \sqrt{2} \bar{F} F (\bar{F} (\theta \psi) + F (\bar{\theta}
 \bar{\psi})) + i (\psi \sigma^m \bar{\psi}) (F \partial_m \bar{\varphi}
 - \bar{F} \partial_m \varphi).
\label{eq:HD}
\end{align}
The component expansion of the $\Lambda$ function for the
model $\Lambda = \lambda - \alpha \partial_m \Phi \partial^m \Phi^{\dagger}$
is calculated using the following expression:
\begin{align}
\partial_m \Phi \partial^m \Phi^{\dagger} =& \
\partial_m \varphi \partial^m \bar{\varphi} + \sqrt{2} (\theta
 \partial_m \psi) \partial^m \bar{\varphi} + \sqrt{2} (\bar{\theta}
 \partial_m \bar{\psi}) \partial^m \varphi
+ \theta^2 \partial_m \bar{\varphi} \partial^m F + \bar{\theta}^2
 \partial_m \varphi \partial^m \bar{F}
\notag \\
& \ + \theta^{\alpha} \bar{\theta}^{\dot{\alpha}}
\left[
\frac{}{}
i (\sigma^p)_{\alpha \dot{\alpha}} (\partial_m \bar{\varphi} \partial_p
 \partial^m \varphi
- \partial_p \partial_m \bar{\varphi} \partial^m \varphi
)
- 2 \partial_m \bar{\psi}_{\dot{\alpha}} \partial^m \psi_{\alpha}
\right]
\notag \\
 & \ + \theta^2 \bar{\theta}^{\dot{\alpha}}
\left[
\frac{i}{\sqrt{2}} (\sigma^p)_{\alpha \dot{\alpha}}
\left(
\partial_m \bar{\varphi} \partial_p \partial^m \psi^{\alpha} -
 \partial_p \partial_m \bar{\varphi} \partial^m \psi^{\alpha}
\right)
- \sqrt{2} \partial_m F \partial^m \bar{\psi}_{\dot{\alpha}}
\right]
\notag \\
 & \ + \bar{\theta}^2 \theta^{\alpha}
\left[
- \frac{i}{\sqrt{2}} (\sigma^p)_{\alpha \dot{\alpha}}
\left(
\partial_m \bar{\psi}^{\dot{\alpha}} \partial_p \partial^m \varphi -
 \partial_p \partial_m \bar{\psi}^{\dot{\alpha}} \partial^m \varphi
\right)
+ \sqrt{2} \partial_m \bar{F} \partial^m \psi_{\alpha}
\right]
\notag \\
& \ + \theta^2 \bar{\theta}^2
\left[
\partial_m F \partial^m \bar{F}
+ \frac{1}{4} \partial_m \bar{\varphi} \Box \partial^m \varphi +
 \frac{1}{4} \Box \partial_m \bar{\varphi} \partial^m \varphi
- \frac{1}{2} \partial_m \partial_p \bar{\varphi} \partial^m \partial^p
 \varphi
\right.
\notag \\
& \qquad \qquad
\left.
+ \frac{i}{2} \partial_m \partial_p \bar{\psi} \bar{\sigma}^p \partial^m
 \psi
- \frac{i}{2} \partial_m \bar{\psi} \bar{\sigma}^p \partial_p \partial^m \psi
\right].
\end{align}

\end{appendix}



\begin{thebibliography}{0}

\bibitem{ORaifeartaigh:1975nky}
  L.~O'Raifeartaigh,
  ``Spontaneous Symmetry Breaking for Chiral Scalar Superfields,''
  Nucl.\ Phys.\ B {\bf 96} (1975) 331.

\bibitem{Fayet:1974jb}
  P.~Fayet and J.~Iliopoulos,
  ``Spontaneously Broken Supergauge Symmetries and Goldstone Spinors,''
  Phys.\ Lett.\  {\bf 51B} (1974) 461.

\bibitem{Witten:1981nf}
  E.~Witten,
  ``Dynamical Breaking of Supersymmetry,''
  Nucl.\ Phys.\ B {\bf 188} (1981) 513.

\bibitem{Witten:1982df}
  E.~Witten,
  ``Constraints on Supersymmetry Breaking,''
  Nucl.\ Phys.\ B {\bf 202} (1982) 253.

\bibitem{Intriligator:2006dd}
  K.~A.~Intriligator, N.~Seiberg and D.~Shih,
  ``Dynamical SUSY breaking in meta-stable vacua,''
  JHEP {\bf 0604} (2006) 021
  [hep-th/0602239].

\bibitem{Intriligator:2007cp}
  K.~A.~Intriligator and N.~Seiberg,
  ``Lectures on Supersymmetry Breaking,''
  Class.\ Quant.\ Grav.\  {\bf 24} (2007) S741
  [hep-ph/0702069].

\bibitem{Ivanov:1975zq}
  E.~A.~Ivanov and V.~I.~Ogievetsky,
  Teor.\ Mat.\ Fiz.\  {\bf 25} (1975) 164.
  doi:10.1007/BF01028947

\bibitem{Brauner:2014aha}
  T.~Brauner and H.~Watanabe,
  ``Spontaneous breaking of spacetime symmetries and the inverse Higgs effect,''
  Phys.\ Rev.\ D {\bf 89} (2014) no.8,  085004
  [arXiv:1401.5596 [hep-ph]].

\bibitem{Kostelecky:1988zi}
  V.~A.~Kostelecky and S.~Samuel,
  ``Spontaneous Breaking of Lorentz Symmetry in String Theory,''
  Phys.\ Rev.\ D {\bf 39} (1989) 683.
%
\bibitem{Bluhm:2004ep}
  R.~Bluhm and V.~A.~Kostelecky,
  ``Spontaneous Lorentz violation, Nambu-Goldstone modes, and gravity,''
  Phys.\ Rev.\ D {\bf 71} (2005) 065008
  [hep-th/0412320].
%
\bibitem{Altschul:2005mu}
  B.~Altschul and V.~A.~Kostelecky,
  ``Spontaneous Lorentz violation and nonpolynomial interactions,''
  Phys.\ Lett.\ B {\bf 628} (2005) 106
  [hep-th/0509068].
%
\bibitem{Kostelecky:2009zr}
  V.~A.~Kostelecky and R.~Potting,
  ``Gravity from spontaneous Lorentz violation,''
  Phys.\ Rev.\ D {\bf 79} (2009) 065018
  [arXiv:0901.0662 [gr-qc]].
%
\bibitem{Bluhm:2007bd}
  R.~Bluhm, S.~H.~Fung and V.~A.~Kostelecky,
  ``Spontaneous Lorentz and Diffeomorphism Violation, Massive Modes, and Gravity,''
  Phys.\ Rev.\ D {\bf 77} (2008) 065020
  [arXiv:0712.4119 [hep-th]].

\bibitem{ArkaniHamed:2003uy}
  N.~Arkani-Hamed, H.~C.~Cheng, M.~A.~Luty and S.~Mukohyama,
  ``Ghost condensation and a consistent infrared modification of gravity,''
  JHEP {\bf 0405} (2004) 074
  [hep-th/0312099].

\bibitem{Aganagic:1996nn}
  M.~Aganagic, C.~Popescu and J.~H.~Schwarz,
  ``Gauge invariant and gauge fixed D-brane actions,''
  Nucl.\ Phys.\ B {\bf 495} (1997) 99
  [hep-th/9612080].

\bibitem{Adawi:1998ta}
  T.~Adawi, M.~Cederwall, U.~Gran, B.~E.~W.~Nilsson and B.~Razaznejad,
  ``Goldstone tensor modes,''
  JHEP {\bf 9902} (1999) 001
  [hep-th/9811145].

\bibitem{Kaplan:1995cp}
  D.~M.~Kaplan and J.~Michelson,
  ``Zero modes for the D = 11 membrane and five-brane,''
  Phys.\ Rev.\ D {\bf 53} (1996) 3474
  [hep-th/9510053].

\bibitem{Simon:2011rw}
  J.~Simon,
  ``Brane Effective Actions, Kappa-Symmetry and Applications,''
  Living Rev.\ Rel.\  {\bf 15} (2012) 3
  [arXiv:1110.2422 [hep-th]].

\bibitem{Bagger:1996wp}
  J.~Bagger and A.~Galperin,
  ``A New Goldstone multiplet for partially broken supersymmetry,''
  Phys.\ Rev.\ D {\bf 55} (1997) 1091
  [hep-th/9608177].

\bibitem{Kallosh:1997aw}
  R.~Kallosh,
  ``Volkov-Akulov theory and D-branes,''
  Lect.\ Notes Phys.\  {\bf 509} (1998) 49
  [hep-th/9705118].

\bibitem{Ivanov:1999fwa}
  E.~Ivanov and S.~Krivonos,
  ``$N=1$ $D=2$ supermembrane in the coset approach,''
  Phys.\ Lett.\ B {\bf 453} (1999) 237
   Erratum: [Phys.\ Lett.\ B {\bf 657} (2007) 269]
   Erratum: [Phys.\ Lett.\ B {\bf 460} (1999) 499]
  [hep-th/9901003].

\bibitem{Clark:2002bh}
  T.~E.~Clark, M.~Nitta and T.~ter Veldhuis,
  ``Brane dynamics from nonlinear realizations,''
  Phys.\ Rev.\ D {\bf 67} (2003) 085026
  [hep-th/0208184].

\bibitem{Horava:2009uw}
  P.~Horava,
  ``Quantum Gravity at a Lifshitz Point,''
  Phys.\ Rev.\ D {\bf 79} (2009) 084008
  [arXiv:0901.3775 [hep-th]].

\bibitem{Sotiriou:2009bx}
  T.~P.~Sotiriou, M.~Visser and S.~Weinfurtner,
  ``Quantum gravity without Lorentz invariance,''
  JHEP {\bf 0910} (2009) 033
  [arXiv:0905.2798 [hep-th]].

\bibitem{Dubovsky:2004sg}
  S.~L.~Dubovsky,
  ``Phases of massive gravity,''
  JHEP {\bf 0410} (2004) 076
  [hep-th/0409124].

\bibitem{Nojiri:2010wj}
  S.~Nojiri and S.~D.~Odintsov,
  ``Unified cosmic history in modified gravity: from F(R) theory to Lorentz non-invariant models,''
  Phys.\ Rept.\  {\bf 505} (2011) 59
  [arXiv:1011.0544 [gr-qc]].

\bibitem{Coleman:1998ti}
  S.~R.~Coleman and S.~L.~Glashow,
  ``High-energy tests of Lorentz invariance,''
  Phys.\ Rev.\ D {\bf 59} (1999) 116008
  [hep-ph/9812418].
%
\bibitem{Kostelecky:2003fs}
  V.~A.~Kostelecky,
  ``Gravity, Lorentz violation, and the standard model,''
  Phys.\ Rev.\ D {\bf 69} (2004) 105009
  [hep-th/0312310].

\bibitem{Colladay:1998fq}
  D.~Colladay and V.~A.~Kostelecky,
  ``Lorentz violating extension of the standard model,''
  Phys.\ Rev.\ D {\bf 58} (1998) 116002
  [hep-ph/9809521].

\bibitem{Berger:2001rm}
  M.~S.~Berger and V.~A.~Kostelecky,
  ``Supersymmetry and Lorentz violation,''
  Phys.\ Rev.\ D {\bf 65} (2002) 091701
  [hep-th/0112243].
%
\bibitem{Berger:2003ay}
  M.~S.~Berger,
  ``Superfield realizations of Lorentz and CPT violation,''
  Phys.\ Rev.\ D {\bf 68} (2003) 115005
  [hep-th/0308036].
%
\bibitem{Bolokhov:2005cj}
  P.~A.~Bolokhov, S.~Groot Nibbelink and M.~Pospelov,
  ``Lorentz violating supersymmetric quantum electrodynamics,''
  Phys.\ Rev.\ D {\bf 72} (2005) 015013
  [hep-ph/0505029].
%
\bibitem{GrootNibbelink:2004za}
  S.~Groot Nibbelink and M.~Pospelov,
  ``Lorentz violation in supersymmetric field theories,''
  Phys.\ Rev.\ Lett.\  {\bf 94} (2005) 081601
  [hep-ph/0404271].
%

\bibitem{Fulde:1964zz}
  P.~Fulde and R.~A.~Ferrell,
  ``Superconductivity in a Strong Spin-Exchange Field,''
  Phys.\ Rev.\  {\bf 135} (1964) A550.

\bibitem{larkin:1964zz}
  A.~I.~Larkin and Y.~N.~Ovchinnikov,
  ``Nonuniform state of superconductors,''
  Zh.\ Eksp.\ Teor.\ Fiz.\  {\bf 47} (1964) 1136
   [Sov.\ Phys.\ JETP {\bf 20} (1965) 762].

\bibitem{Nakano:2004cd}
  E.~Nakano and T.~Tatsumi,
  ``Chiral symmetry and density wave in quark matter,''
  Phys.\ Rev.\ D {\bf 71}, 114006 (2005)
  [hep-ph/0411350].

\bibitem{Nickel:2009ke}
  D.~Nickel,
  ``How many phases meet at the chiral critical point?,''
  Phys.\ Rev.\ Lett.\  {\bf 103}, 072301 (2009)
  [arXiv:0902.1778 [hep-ph]];
  D.~Nickel,
  ``Inhomogeneous phases in the Nambu-Jona-Lasino and quark-meson model,''
  Phys.\ Rev.\ D {\bf 80}, 074025 (2009)
  [arXiv:0906.5295 [hep-ph]].

\bibitem{Buballa:2014tba}
  M.~Buballa and S.~Carignano,
  ``Inhomogeneous chiral condensates,''
  Prog.\ Part.\ Nucl.\ Phys.\  {\bf 81} (2015) 39
  [arXiv:1406.1367 [hep-ph]].



\bibitem{Nitta:2017mgk}
  M.~Nitta, S.~Sasaki and R.~Yokokura,
  ``Spatially Modulated Vacua in Relativistic Field Theories,''
  arXiv:1706.02938 [hep-th].

\bibitem{Antoniadis:2007xc}
  I.~Antoniadis, E.~Dudas and D.~M.~Ghilencea,
  ``Supersymmetric Models with Higher Dimensional Operators,''
  JHEP {\bf 0803} (2008) 045
  [arXiv:0708.0383 [hep-th]];
  E.~Dudas and D.~M.~Ghilencea,
  ``Effective operators in SUSY, superfield constraints and searches for a UV completion,''
  JHEP {\bf 1506}, 124 (2015)
  [arXiv:1503.08319 [hep-th]].


\bibitem{Ostrogradski}
  M.~Ostrogradski, ``Memoires sur les equations differentielles relatives au probleme des
  isoperimetres,''
  Mem. \ Ac. \ St. Petersbourg VI (1850) 385.


\bibitem{Fujimori:2016udq}
  T.~Fujimori, M.~Nitta and Y.~Yamada,
  ``Ghostbusters in higher derivative supersymmetric theories: who is afraid of propagating auxiliary fields?,''
  JHEP {\bf 1609}, 106 (2016)
  [arXiv:1608.01843 [hep-th]].


\bibitem{Gates:1995fx}
  S.~J.~Gates, Jr.,
  ``Why auxiliary fields matter: The Strange case of the 4-D, N=1 supersymmetric QCD effective action,''
  Phys.\ Lett.\ B {\bf 365}, 132 (1996)
  [hep-th/9508153];
  ``Why auxiliary fields matter: The strange case of the 4-D, N=1 supersymmetric QCD effective action. 2.,''
  Nucl.\ Phys.\ B {\bf 485}, 145 (1997)
  [hep-th/9606109].


\bibitem{Nemeschansky:1984cd}
  D.~Nemeschansky and R.~Rohm,
  ``Anomaly Constraints On Supersymmetric Effective Lagrangians,''
  Nucl.\ Phys.\ B {\bf 249}, 157 (1985);
  M.~Nitta,
  ``A Note on supersymmetric WZW term in four dimensions,''
  Mod.\ Phys.\ Lett.\ A {\bf 15}, 2327 (2000)
  [hep-th/0101166].


\bibitem{Bergshoeff:1984wb}
  E.~A.~Bergshoeff, R.~I.~Nepomechie and H.~J.~Schnitzer,
  ``Supersymmetric Skyrmions in Four-dimensions,''
  Nucl.\ Phys.\ B {\bf 249} (1985) 93;
  L.~Freyhult,
  ``The Supersymmetric extension of the Faddeev model,''
  Nucl.\ Phys.\ B {\bf 681} (2004) 65
  [hep-th/0310261].



\bibitem{Buchbinder:1994iw}
  I.~L.~Buchbinder, S.~Kuzenko and Z.~Yarevskaya,
  ``Supersymmetric effective potential: Superfield approach,''
  Nucl.\ Phys.\ B {\bf 411}, 665 (1994);
  I.~L.~Buchbinder, S.~M.~Kuzenko and A.~Y.~Petrov,
  ``Superfield chiral effective potential,''
  Phys.\ Lett.\ B {\bf 321} (1994) 372;
  A.~T.~Banin, I.~L.~Buchbinder and N.~G.~Pletnev,
  ``On quantum properties of the four-dimensional generic chiral superfield model,''
  Phys.\ Rev.\ D {\bf 74}, 045010 (2006)
  [hep-th/0606242].

\bibitem{Gomes:2009ev}
  M.~Gomes, J.~R.~Nascimento, A.~Y.~Petrov and A.~J.~da Silva,
  ``On the effective potential in higher-derivative superfield theories,''
  Phys.\ Lett.\ B {\bf 682} (2009) 229
  [arXiv:0908.0900 [hep-th]];
  F.~S.~Gama, M.~Gomes, J.~R.~Nascimento, A.~Y.~Petrov and A.~J.~da Silva,
  ``On the higher-derivative supersymmetric gauge theory,''
  Phys.\ Rev.\ D {\bf 84} (2011) 045001
  [arXiv:1101.0724 [hep-th]].



\bibitem{Gates:2000rp}
  S.~J.~Gates, Jr., M.~T.~Grisaru, M.~E.~Knutt and S.~Penati,
  ``The Superspace WZNW action for 4-D, N=1 supersymmetric QCD,''
  Phys.\ Lett.\ B {\bf 503}, 349 (2001)
  [hep-ph/0012301];
  S.~J.~Gates, Jr., M.~T.~Grisaru, M.~E.~Knutt, S.~Penati and H.~Suzuki,
  ``Supersymmetric gauge anomaly with general homotopic paths,''
  Nucl.\ Phys.\ B {\bf 596}, 315 (2001)
  [hep-th/0009192];
  S.~J.~Gates, Jr., M.~T.~Grisaru and S.~Penati,
  ``Holomorphy, minimal homotopy and the 4-D, N=1 supersymmetric Bardeen-Gross-Jackiw anomaly,''
  Phys.\ Lett.\ B {\bf 481}, 397 (2000)
  [hep-th/0002045].

\bibitem{Rocek:1997hi}
  M.~Rocek and A.~A.~Tseytlin,
  ``Partial breaking of global D = 4 supersymmetry, constrained superfields, and three-brane actions,''
  Phys.\ Rev.\ D {\bf 59} (1999) 106001
  [hep-th/9811232].

\bibitem{Farakos:2012qu}
  F.~Farakos and A.~Kehagias,
  ``Emerging Potentials in Higher-Derivative Gauged Chiral Models Coupled to N=1 Supergravity,''
  JHEP {\bf 1211} (2012) 077
  [arXiv:1207.4767 [hep-th]];
  F.~Farakos, O.~Hulik, P.~Ko\v{c}i and R.~von Unge,
  ``Non-minimal scalar multiplets, supersymmetry breaking and dualities,''
  JHEP {\bf 1509}, 177 (2015)
  [arXiv:1507.01885 [hep-th]].

\bibitem{Adam:2011gc}
  C.~Adam, J.~M.~Queiruga, J.~Sanchez-Guillen and A.~Wereszczynski,
  ``Supersymmetric K field theories and defect structures,''
  Phys.\ Rev.\ D {\bf 84} (2011) 065032
  [arXiv:1107.4370 [hep-th]];
  C.~Adam, J.~M.~Queiruga, J.~Sanchez-Guillen and A.~Wereszczynski,
  ``N=1 supersymmetric extension of the baby Skyrme model,''
  Phys.\ Rev.\ D {\bf 84}, 025008 (2011)
  [arXiv:1105.1168 [hep-th]];
  C.~Adam, J.~M.~Queiruga, J.~Sanchez-Guillen and A.~Wereszczynski,
  ``Extended Supersymmetry and BPS solutions in baby Skyrme models,''
  JHEP {\bf 1305} (2013) 108
  [arXiv:1304.0774 [hep-th]];
  S.~Bolognesi and W.~Zakrzewski,
  Phys.\ Rev.\ D {\bf 91}, no. 4, 045034 (2015)
  [arXiv:1407.3140 [hep-th]];
  J.~M.~Queiruga,
  ``Skyrme-like models and supersymmetry in 3+1 dimensions,''
  Phys.\ Rev.\ D {\bf 92}, no. 10, 105012 (2015)
  [arXiv:1508.06692 [hep-th]];
  J.~M.~Queiruga,
  ``Supersymmetric galileons and auxiliary fields in 2+1 dimensions,''
  Phys.\ Rev.\ D {\bf 95}, no. 12, 125001 (2017)
  [arXiv:1612.04727 [hep-th]].

\bibitem{Gudnason:2015ryh}
  S.~B.~Gudnason, M.~Nitta and S.~Sasaki,
  ``A supersymmetric Skyrme model,''
  JHEP {\bf 1602} (2016) 074
  [arXiv:1512.07557 [hep-th]];
  S.~B.~Gudnason, M.~Nitta and S.~Sasaki,
  ``Topological solitons in the supersymmetric Skyrme model,''
  JHEP {\bf 1701} (2017) 014
  [arXiv:1608.03526 [hep-th]].

\bibitem{Sasaki:2012ka}
  S.~Sasaki, M.~Yamaguchi and D.~Yokoyama,
  ``Supersymmetric DBI inflation,''
  Phys.\ Lett.\ B {\bf 718} (2012) 1
  [arXiv:1205.1353 [hep-th]].

\bibitem{Nitta:2014pwa}
  M.~Nitta and S.~Sasaki,
  ``BPS States in Supersymmetric Chiral Models with Higher Derivative Terms,''
  Phys.\ Rev.\ D {\bf 90}, no. 10, 105001 (2014)
  [arXiv:1406.7647 [hep-th]];
  M.~Nitta and S.~Sasaki,
  ``Classifying BPS States in Supersymmetric Gauge Theories Coupled to Higher Derivative Chiral Models,''
  Phys.\ Rev.\ D {\bf 91} (2015) 125025
  [arXiv:1504.08123 [hep-th]].

\bibitem{Nitta:2014fca}
  M.~Nitta and S.~Sasaki,
  ``Higher Derivative Corrections to Manifestly Supersymmetric Nonlinear Realizations,''
  Phys.\ Rev.\ D {\bf 90} (2014) no.10,  105002
  [arXiv:1408.4210 [hep-th]].

\bibitem{Khoury:2010gb}
  J.~Khoury, J.~-L.~Lehners and B.~Ovrut,
  ``Supersymmetric P(X,$\phi$) and the Ghost Condensate,''
  Phys.\ Rev.\ D {\bf 83} (2011) 125031
  [arXiv:1012.3748 [hep-th]];
  M.~Koehn, J.~L.~Lehners and B.~Ovrut,
  ``Ghost condensate in $N=1$ supergravity,''
  Phys.\ Rev.\ D {\bf 87} (2013) no.6,  065022
  [arXiv:1212.2185 [hep-th]];
  M.~Koehn, J.~L.~Lehners and B.~A.~Ovrut,
  ``Higher-Derivative Chiral Superfield Actions Coupled to N=1 Supergravity,''
  Phys.\ Rev.\ D {\bf 86} (2012) 085019
  [arXiv:1207.3798 [hep-th]].

\bibitem{Wess:1992cp}
  J.~Wess and J.~Bagger,
  ``Supersymmetry and supergravity,''
  Princeton, USA: Univ. Pr. (1992) 259 p

\bibitem{Low:2001bw}
  I.~Low and A.~V.~Manohar,
  ``Spontaneously broken space-time symmetries and Goldstone's theorem,''
  Phys.\ Rev.\ Lett.\  {\bf 88} (2002) 101602
  [hep-th/0110285].

\bibitem{Kugo:2017qma}
  T.~Kugo,
  ``Spontaneous Supersymmetry Breaking, Negative Metric and Vacuum Energy,''
  arXiv:1703.00600 [hep-th].

\bibitem{Ohta:1981bi}
  N.~Ohta,
  ``Explicit Soft Versus Spontaneous Breakings of Supersymmetry,''
  Phys.\ Lett.\  {\bf 112B} (1982) 215.

\bibitem{Ohta:1982ys}
  N.~Ohta and Y.~Fujii,
  ``Dipole Mechanism of Spontaneous Breaking of Supersymmetry,''
  Nucl.\ Phys.\ B {\bf 202} (1982) 477.

\bibitem{Cheng:2006us}
  H.~C.~Cheng, M.~A.~Luty, S.~Mukohyama and J.~Thaler,
  ``Spontaneous Lorentz breaking at high energies,''
  JHEP {\bf 0605} (2006) 076
  [hep-th/0603010].

\bibitem{Kojo:2011cn}
  T.~Kojo, Y.~Hidaka, K.~Fukushima, L.~D.~McLerran and R.~D.~Pisarski,
  ``Interweaving Chiral Spirals,''
  Nucl.\ Phys.\ A {\bf 875}, 94 (2012)
  [arXiv:1107.2124 [hep-ph]].

\bibitem{Hayata:2013sea}
  T.~Hayata, Y.~Hidaka and A.~Yamamoto,
  ``Temporal chiral spiral in QCD in the presence of strong magnetic fields,''
  Phys.\ Rev.\ D {\bf 89}, no. 8, 085011 (2014)
  [arXiv:1309.0012 [hep-ph]].




\end{thebibliography}
\end{document}